\documentclass[review]{elsarticle}

\usepackage{lineno,hyperref}

\usepackage{graphicx}
\graphicspath{ {./figs/} }
\usepackage[dvipsnames,table,xcdraw]{xcolor}
\usepackage{amsmath,amssymb,mathrsfs}
\usepackage{multirow}
\usepackage[utf8]{inputenc}
\usepackage{url}

\usepackage{tikz}
\usetikzlibrary{shadows,arrows,shapes.geometric} 
\usetikzlibrary{shapes,fit,backgrounds,patterns,automata}
\usetikzlibrary{plotmarks}

\usepackage{lscape}

\usepackage{pdflscape}

\usepackage{xfrac}

\usepackage{amsmath}

\journal{Computer Physics Communications}

\begin{document}

\begin{frontmatter}

\title{Cache Blocking for Flux Reconstruction:\\\ Extension to Navier-Stokes Equations and Anti-aliasing}


\author[mymainaddress]{Semih Akkurt\corref{mycorrespondingauthor}}

\author[mysecondaryaddress]{Freddie Witherden}

\author[mymainaddress]{Peter Vincent}

\address[mymainaddress]{\scriptsize{Department of Aeronautics, Imperial College London, SW7 2AZ, United Kingdom}}
\address[mysecondaryaddress]{\scriptsize{Department of Ocean Engineering, Texas A\&M University, College Station, TX 77843, USA}}

\begin{abstract}
In this article, cache blocking is implemented for the Navier Stokes equations with anti-aliasing support on mixed grids in PyFR for CPUs.
In particular, cache blocking is used as an alternative to kernel fusion to eliminate unnecessary data movements between kernels at the main memory level.
Specifically, kernels that exchange data are grouped together, and these groups are then executed on small sub-regions of the domain that fit in per-core private data cache.
Additionally, cache blocking is also used to efficiently implement a tensor product factorisation of the interpolation operators associated with anti-aliasing.
By using cache blocking, the intermediate results between application of the sparse factors are stored in per-core private data cache, and a significant amount of data movement from main memory is avoided.
In order to assess the performance gains a theoretical model is developed, and the implementation is benchmarked using a compressible 3D Taylor-Green vortex test case on both hexahedral and prismatic grids, with third- and forth-order solution polynomials. 
The expected performance gains based on the theoretical model range from 1.99 to 2.62, and the speedups obtained in practice range from 1.67 to 3.67 compared to PyFR v1.11.0.
\end{abstract}


\end{frontmatter}


\section{Introduction}

Achieving a target accuracy level with as little computational cost as possible has always been an important goal in Computational Fluid Dynamics (CFD).
To this end, various numerical discretizations have been developed and improved over the years. 
For example, compact high-order discretizations for unstructured grids, such as Discontinuous Galerkin (DG), Spectral Difference (SD), and Flux Reconstruction (FR) schemes have been developed to enable accurate scale-resolving simulations of turbulent flow in a tractable cost envelope. 
However, scale resolving simulations of various industry-relevant high Reynolds number flows are still beyond the reach of high-order methods even when using national laboratory scale supercomputers.
This has led researchers to investigate ways to utilize available computational power more efficiently. 
Changing the arithmetic intensity of kernels can be a way to achieve this.
For example, an alternative mathematical form of a given operation may  require fewer FLOPs and this may lead to a better arithmetic intensity for a particular hardware.
Another example is kernel fusion, which reduces the data movements between processor and memory, and therefore increases the arithmetic intensity.
Depending on the arithmetic intensity profile of the kernels and the target hardware, these methods can be used to achieve better performance.

In recent work, the authors implemented a cache blocking strategy as an alternative to kernel fusion to reduce bandwidth requirement for an FR code for Euler equations without anti-aliasing on hexahedral meshes \cite{Akkurt}.
Cache blocking reduces main memory accesses by using L2 cache as a buffer storage for intermediate and temporary data in a set of kernels that exchange data, which we will refer as a kernel group in the remainder of this article. 
As long as the L2 cache is big enough for storing the temporary data, cache blocking has all the benefits of kernel fusion in terms of bandwidth reduction without any downsides \cite{Akkurt}.
Furthermore, the cache blocking approach is easy to implement, and changing the configuration of kernel groups or adding new kernels into the existing kernel groups is simple and straightforward.
Here, we extend this previous work to support Navier-Stokes equations with anti-aliasing on mixed element meshes.
Extension to Navier-Stokes solver enables real-world viscous simulations and it is necessary to use anti-aliasing in order to remain numerically stable for certain physical simulations such as turbulent flows that are marginally resolved \cite{Park17}.
Additionally, supporting mixed element meshes provides flexibility for complex geometries.

There are three important aspects to consider for the aforementioned extensions.
First, compared to an Euler solver, a Navier-Stokes solver has more kernels and this introduces more constraints on how they can be grouped together. 
Next, a direct implementation of anti-aliasing results in additional dense operators as well as changing many of the existing sparse operators in the Navier-Stokes solver into dense operators.
This results in FLOP limited kernels due to the sufficiently high arithmetic intensity associated with the dense operators.
However, an alternative method for implementing anti-aliasing is using sum factorisation, which was initially proposed by Orszag \cite{ORSZAG198070} for spectral methods and later adapted and used for spectral/$hp$ element methods by Sherwin and Karniadakis \cite{Sherwin95}. 
A detailed analysis of sum factorisation carried out by Cantwell \cite{Cantwell11} shows that sum factorisation can be effective only at $p=7$ and beyond.
Further studies also confirmed a similar trend \cite{Moxey16, Bolis14}.
Świrydowicz et al. \cite{Warburton19} employed a slightly different strategy which factors the tensor product operators into sparse components by doing 1D interpolations in every dimension one-by-one.
Employing such a tensor product factorisation strategy naively for these dense operators where each factor is applied sequentially to the entire data set without cache blocking results in significant increase in the bandwidth requirement, and makes the operations bandwidth bound rather than FLOP bound.
Kernel fusion is a solution to this, but it requires not only a complex implementation, but also makes the code architecture specific.
Therefore, we propose using the cache blocking strategy in conjunction with the tensor product factorisation in order to eliminate the excess bandwidth requirement. 
In short, the methodology we propose for anti-aliasing effectively trades of FLOP requirements with bandwidth requirements by using a tensor product factorisation strategy, and then eliminates the excess bandwidth requirement by employing a cache blocking strategy.
As a result, anti-aliasing kernels require significantly less FLOPs while requiring similar or even reduced access to main memory.
Finally, we enable mixed element support including hexahedra, prisms, tetrahedra and pyramids where tensor product factorisation is only used for hexahedral and prism elements.

The paper is structured as follows.
The FR formulation for the Navier-Stokes equations including anti-aliasing is given in Section \ref{sec:FR} alongside with the kernels that implement these mathematical operations.
Next, current kernel execution order and the data movement requirements are presented in Section \ref{sec:kex}
Anti-aliasing and its effects in terms of computational cost is discussed in Section \ref{sec:AA}.
Then, an efficient implementation of anti-aliasing in FR within the context of cache blocking is described in Section \ref{sec:cleverthing}.
A kernel grouping strategy for the Navier-Stokes equations including anti-aliasing support is given alongside the speedups obtained in practice in Section \ref{sec:grouping}.
Finally, conclusions are discussed in Section \ref{sec:conclusion}.

\section{Navier-Stokes Formulation and Kernels}\label{sec:FR}
Flux Reconstruction (FR) was first developed by Huynh in 2005 \cite{huynh}.
An overview of the FR approach for solving the Navier-Stokes equations including anti-aliasing is presented in this section.

\subsection{Formulation}\label{sec:FRfor}
The Navier-Stokes equations can be written in conservative form as,
\begin{equation}\label{eq:gov}
    \frac{\partial u_{\alpha}}{\partial t}
    +
    \nabla \cdot \mathbf{f}_{\alpha}
    =
    0,
\end{equation}
where $\alpha$ is the field variable index, $u_\alpha=u_\alpha(\mathbf{x},t)$ are the conservative field variables thus,
\begin{equation}\label{eq:ucons}
u
=
\begin{bmatrix}
\rho \\
\rho v_x \\
\rho v_y \\
\rho v_z \\
E
\end{bmatrix},
\end{equation} 
and $\mathbf{f}=\mathbf{f}(u, \nabla u)$ is the flux term that incorporates inviscid and viscous fluxes thus $\mathbf{f}=\mathbf{f}^e-\mathbf{f}^v$. 
First, the inviscid flux, $\mathbf{f}^e$, is defined as
\begin{equation}\label{eq:fluxe}
\mathbf{f}^e(u)
=
\begin{bmatrix}
\rho v_x \\
\rho v_x^2 + p \\
\rho v_x v_y \\
\rho v_x v_z \\
v_x (E+p)
\end{bmatrix}
\mathbf{i}
+
\begin{bmatrix}
\rho v_y \\
\rho v_y v_x \\
\rho v_y^2 + p \\
\rho v_y v_z \\
v_y (E+p)
\end{bmatrix}
\mathbf{j}
+
\begin{bmatrix}
\rho v_z \\
\rho v_z v_x \\
\rho v_z v_y \\
\rho v_z^2 + p \\
v_z (E+p)
\end{bmatrix}
\mathbf{k},
\end{equation}
where $\rho$ is density, $v_x, v_y, v_z$ are velocity, $p$ is pressure, $E$ is total energy per unit volume, and $\mathbf{i, j, k}$ are orthogonal unit vectors.
The relation between the pressure and the total energy for a perfect gas is,
\begin{equation}
    E = \frac{p}{\gamma -1} + \frac{1}{2}\rho(v_x^2+v_y^2+v_z^2),
\end{equation}
where $\gamma$ is the specific heat ratio, $c_p/c_v$, $c_p$ is the specific heat at constant pressure, and $c_v$ is the specific heat at constant volume.

Next, the viscous flux, $\mathbf{f}^v$ is given as
\begin{equation}\label{eq:fluxv}
\mathbf{f}^v(u, \Delta u)
=
\begin{bmatrix}
0 \\
\mathcal{T}_{xx} \\
\mathcal{T}_{xy} \\
\mathcal{T}_{xz} \\
v_i \mathcal{T}_{ix} + \phi_x
\end{bmatrix}
\mathbf{i}
+
\begin{bmatrix}
0 \\
\mathcal{T}_{yx} \\
\mathcal{T}_{yy} \\
\mathcal{T}_{yz} \\
v_i \mathcal{T}_{iy} + \phi_y
\end{bmatrix}
\mathbf{j}
+
\begin{bmatrix}
0 \\
\mathcal{T}_{zx} \\
\mathcal{T}_{zy} \\
\mathcal{T}_{zz} \\
v_i \mathcal{T}_{iz} + \phi_z
\end{bmatrix}
\mathbf{k},
\end{equation}
where $\mathcal{T}$ is the stress tensor and defined as, 
\begin{equation}
\mathcal{T} = \mu(\partial_iv_j + \partial_jv_i) - \frac{2}{3}\mu \delta_{ij}\nabla \cdot \mathbf{v},
\end{equation}
and $\mathbf{\phi}$ is the heat flux defined as,
\begin{equation}
\mathbf{\phi} = \mu \frac{c_p}{P_r}\nabla T.
\end{equation}
The ideal gas law relates the temperature $T$ with pressure and density as
\begin{equation}
T = \frac{1}{c_v}\frac{1}{1-\gamma}\frac{p}{\rho}.
\end{equation}

With the flux term defined, $\nabla u$ can be substituted by $\mathbf{q}$ so that Equation \ref{eq:gov} can be rewritten as a first order system thus,
\begin{equation}\label{eq:govq}
\begin{split}
\frac{\partial u}{\partial t} + \nabla \cdot \mathbf{f}_\alpha(u,\mathbf{q}) &= 0,\\
\mathbf{q}_\alpha - \nabla u_\alpha &= 0.
\end{split}
\end{equation}

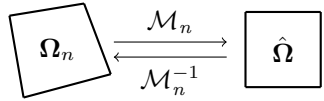
\begin{figure}[!h]
    \centering
\begin{tikzpicture}
\begin{scope}[rotate around={15:(0.5,0.5)}]
\draw [thick] (-0.1,0) to (0,1.1);
\draw [thick] (0,1.1) to (1,1);
\draw [thick] (1,1) to (1.1,0);
\draw [thick] (1.1,0) to (-0.1,0);
\end{scope}
\node at (0.5, 0.5) {$\mathbf{\Omega}_n$}; 
\path [draw,->] (1.25,0.6) -- node [midway,above] {$\mathbf{\mathcal{M}}_\mathit{n}$} (2.75,0.6);

\path [draw,->] (2.75,0.4) -- node [midway,below] {$\mathbf{\mathcal{M}}_\mathit{n}^{-1}$} (1.25,0.4);
\draw [thick] (3,0) to (3,1);
\draw [thick] (3,1) to (4,1);
\draw [thick] (4,1) to (4,0);
\draw [thick] (4,0) to (3,0);
\node at (3.5,0.5) {$\hat{\mathbf{\Omega}}$};
\end{tikzpicture}
    \caption{Mapping function from physical space to reference space for a quadrilateral element.}
    \label{fig:mapping}
\end{figure}

When applying FR to solve the Navier-Stokes equations in a domain $\mathbf{\Omega}$, the first step is to tessellate the domain with non-overlapping, conforming elements thus
\begin{equation}
    \mathbf{\Omega} = \bigcup_{e\in \mathcal{E}} \mathbf{\Omega}_{e}, \qquad
    \mathbf{\Omega}_e = \bigcup_{n=1}^{|\mathbf{\Omega}_e|} \mathbf{\Omega}_{en}, \qquad
    \bigcap_{e\in \mathcal{E}}\bigcap_{n=1}^{|\mathbf{\Omega}_e|}\mathbf{\Omega}_{en} = \varnothing.
\end{equation}
Where $e$ is a particular element type in the list of available elements $\mathcal{E}$, therefore $\Omega_e$ refers to all elements of type $e$, and $|\Omega_e|$ is the number of elements of type $e$.
Each element $\mathbf{\Omega}_{en}$ is then mapped to a reference element $\hat{\mathbf{\Omega}}_e$ via a mapping function $\mathbf{\mathcal{M}}_\mathit{en}$ defined as
\begin{equation*}
    \mathbf{x} = \mathbf{\mathcal{M}}_\mathit{en}(\Tilde{\mathbf{x}}), \qquad
    \mathbf{\Tilde{x}} = \mathbf{\mathcal{M}}_\mathit{en}^{-1}(\mathbf{x}),
\end{equation*}
and this is illustrated for a quadrilateral element in Figure \ref{fig:mapping}.

Geometric Jacobian matrices can be defined from the mapping functions thus
\begin{alignat*}{4}
    \mathbf{J}_\mathit{en} &= J_\mathit{en\!i\!j} &&= \frac{\partial \mathcal{M}_\mathit{en\!i}}{\partial \Tilde{x}_j}, \qquad
    && J_\mathit{en} &&= \det \mathbf{J}_\mathit{en}, \\
    \mathbf{J}^{-1}_\mathit{en} &= J^{-1}_\mathit{en\!i\!j} &&= \frac{\partial \mathcal{M}^{-1}_\mathit{en\!i}}{\partial \Tilde{x}_j}, \qquad
    && J^{-1}_\mathit{en} &&= \det \mathbf{J}^{-1}_\mathit{en}=\frac{1}{J_\mathit{en}}.
\end{alignat*}

The definitions above will be used to transform quantities to and from reference element space.
In order to transform Equation \ref{eq:govq} into reference space, transformed solution $ \Tilde{u}_\mathit{n\alpha} $, transformed flux $\Tilde{\mathbf{f}}_\mathit{n\alpha} $, and transformed gradient $\Tilde{\mathbf{q}}_{\mathit{en\alpha}} $ are defined thus 
\begin{subequations}
\begin{alignat}{2}
    \Tilde{u}_\mathit{en\alpha} 
    &= \Tilde{u}_\mathit{en\alpha}(\Tilde{\mathbf{x}},t)
    &&= J_\mathit{en}(\Tilde{\mathbf{x}})u_\mathit{en\alpha}(\mathcal{M}_\mathit{en}(\Tilde{\mathbf{x}}),t),\label{eq:transu} \\
    \Tilde{\mathbf{f}}_\mathit{en\alpha} 
    &= \Tilde{\mathbf{f}}_\mathit{en\alpha}(\Tilde{\mathbf{x}},t)
    &&= J_\mathit{en}(\Tilde{\mathbf{x}})\mathbf{J}^{-1}_\mathit{en}(\mathcal{M}_\mathit{en}(\Tilde{\mathbf{x}}))\mathbf{f}_\mathit{en\alpha}(\mathcal{M}_\mathit{en}(\Tilde{\mathbf{x}}),t),\label{eq:transf} \\
    \Tilde{\mathbf{q}}_{\mathit{en\alpha}} 
    &= \Tilde{\mathbf{q}}_\mathit{en\alpha}(\Tilde{\mathbf{x}},t)
    &&= \mathbf{J}^{T}_\mathit{en}(\Tilde{\mathbf{x}})\mathbf{q}_\mathit{en\alpha}(\mathcal{M}_\mathit{en}(\Tilde{\mathbf{x}}),t),\label{eq:transq}
\end{alignat}
\label{eq:transall}
\end{subequations}
and for a more compact definition let $\Tilde{\nabla} = \partial / \partial \Tilde{x}_i$.
Then Equation \ref{eq:govq} can be written as, 
\begin{equation}\label{eq:govqtrans}
\begin{split}
    \frac{\partial u_{en\alpha}}{\partial t}
    +
    J^{-1}_\mathit{en}\Tilde{\nabla} \cdot \Tilde{\mathbf{f}}_{en\alpha}
    &=
    0,\\
    \Tilde{\mathbf{q}}_{en\alpha} - \Tilde{\nabla} u_{en\alpha} &= 0.
\end{split}
\end{equation}

Now that the equation in the discretized domain is defined, the next step is to define set of solution points inside the elements and flux points at element interfaces.
FR method represents the solution inside the elements by using a polynomial of order $p$.
The solution across the element interfaces is allowed to be discontinuous.
Selection of the solution and flux point sets affects the properties of the numerical scheme as indicated in various publications such as \cite{WITHERDEN2021113014,WITHERDEN20151232qq,tetrasolp}. 
The formulation given in this section is valid for any solution and flux point set.

\begin{figure}[!h]
    \centering
\begin{tikzpicture}[thick,scale=2, x = {(0.9cm,0.1cm)}, y={(0.2cm,0.8cm)}, z={(0.5cm,-0.5cm)},]
\draw (-1, -1, 0) -- (-1, 1, 0) -- (1, 1, 0) -- (1, -1, 0) -- (-1, -1, 0);

\foreach \x in {-0.7746,0,0.7746}{
\foreach \y in {-0.7746,0,0.7746}{
\filldraw[color=red] (\x, \y, 0) circle (1pt);
}
\node[fill=blue,regular polygon, regular polygon sides=3,inner sep=1pt] at (-1, \x, 0) {};
\node[fill=blue,regular polygon, regular polygon sides=3,inner sep=1pt] at (1, \x, 0) {};

\node[fill=blue,regular polygon, regular polygon sides=3,inner sep=1pt] at (\x, 1, 0) {};
\node[fill=blue,regular polygon, regular polygon sides=3,inner sep=1pt] at (\x, -1, 0) {};

\node[fill=blue,regular polygon, regular polygon sides=3,inner sep=1pt] at (\x+2, -1, 0) {};
\node[fill=blue,regular polygon, regular polygon sides=3,inner sep=1pt] at (\x+2, -1*\x, 0) {};
}

\draw (1, -1, 0) -- (1, 1, 0) -- (3, -1, 0) -- (1, -1, 0);

\filldraw[color=red] (2-0.81684757298044, 0.63369514596088, 0) circle (1pt);
\filldraw[color=red] (2+0.63369514596088, -0.81684757298044, 0) circle (1pt);
\filldraw[color=red] (2-0.81684757298044, -0.81684757298044, 0) circle (1pt);
\filldraw[color=red] (2-0.108103018168072, -0.783793963663856, 0) circle (1pt);
\filldraw[color=red] (2-0.783793963663856, -0.108103018168072, 0) circle (1pt);
\filldraw[color=red] (2-0.108103018168072, -0.108103018168072, 0) circle (1pt);
\end{tikzpicture}
    \caption{Solution points and flux points for a quadrilateral and a triangular elements.}
    \label{fig:ufn}
\end{figure}
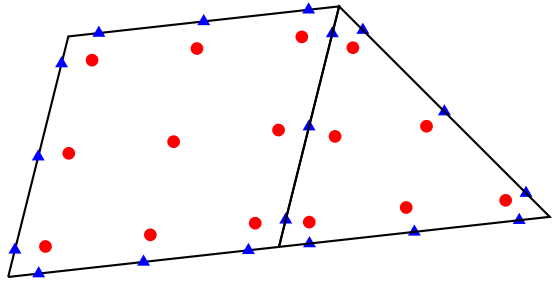

First, let $\mathbf{\Tilde{x}}_{e\zeta}^{(u)}$ be the set of solution points for each reference element $e \in \mathcal{E}$.
$\zeta$ is the solution point index in an element and satisfies $0 \leqslant \zeta < N_e^{(u)}$, where $N_e^{(u)}$ is the number of solution points in element type $e$. 
Now a nodal basis set $\ell_{e\zeta}^{(u)}(\mathbf{\Tilde{x}})$ can be defined where the nodal basis polynomial $\ell_{e\zeta}^{(u)}$ satisfies $\ell_{e\zeta}^{(u)}(\mathbf{\Tilde{x}}_{e\sigma}^{(u)}) = \delta_{\zeta \sigma}$. 
Next, a set of flux points $\mathbf{\Tilde{x}}_{e\zeta}^{(f)}$ on $\partial \hat{\mathbf{\Omega}}_e$ is also defined, where $0 \leqslant \zeta < N_e^{(f)}$ and $N_e^{(f)}$ is the number of flux points at the interfaces of the element type $e$.
These flux points are constrained such that all paired flux points across interfaces match in terms of global coordinates, $\mathbf{\mathcal{M}}_\mathit{en}(\Tilde{\mathbf{x}}_{e\zeta}^{(f)}) = \mathbf{\mathcal{M}}_\mathit{e'n'}(\Tilde{\mathbf{x}}_{e'\zeta'}^{(f)})$, where $'$ denotes the corresponding interface from the neighbouring element. 
Furthermore, there is an associated outward-pointing normal vector for each flux point given by $\hat{\Tilde{\mathbf{n}}}_{e\zeta}^{(f)}$.
The solution points, flux points, 
and an interface between a quadrilateral element and triangular element are all demonstrated in Figure \ref{fig:ufn}.

The first step in the FR method is obtaining the discontinuous solution at the flux points $ u_{\mathit{e\sigma n\alpha}}^{(f)}$ from the solution state at the solution points $ u^{(u)}_\mathit{e\zeta n\alpha}$ as
\begin{equation}\label{eq:FR-disu}
    u^{(f)}_\mathit{e\sigma n\alpha}
    =
    u^{(u)}_\mathit{e\zeta n\alpha}\ell^{(u)}_\mathit{e\zeta}
      (\Tilde{\mathbf{x}}^{(f)}_\mathit{e\sigma})
\end{equation}

Then, the discontinuous solution at the interfaces can be used to obtain a common solution via the LDG method \cite{hesthaven2007nodal,CASTONGUAY2013400} which is defined as
\begin{equation}
\mathfrak{C}(u_L, u_R) = \big( \sfrac{1}{2} - \beta \big)u_L 
                       + \big( \sfrac{1}{2} + \beta \big)u_R.
\end{equation}
It is important to note that $\mathfrak{C}(u_L, u_R)$ is not in general equal to $\mathfrak{C}(u_R, u_L)$.
Therefore it is important to visit all the flux point pairs in the domain only once and assign the common value to both points.
Consequently, the following equalities hold
\begin{equation}\label{eq:FR-conu}
\mathfrak{C}_\alpha (u^{(f)}_\mathit{e\sigma\!n\!\alpha}, u^{(f)}_{\widetilde{\mathit{e\sigma\!n}}\!\alpha})
=
\mathfrak{C}_\alpha u^{(f)}_{\mathit{e\sigma\!n\!\alpha}^{}}
= \mathfrak{C}_\alpha u^{(f)}_{\widetilde{\mathit{e\sigma\!n}}\!\alpha}.
\end{equation}

A correction procedure is then used to obtain a $C^0$ continuous solution across elements.
For this purpose, there is a correction function associated with each flux point in the reference element.
$ \mathbf{g}^\mathit{(f)}_\mathit{e\zeta} $ is the vector correction function which satisfies $
    \hat{\Tilde{\mathbf{n}}}^\mathit{(f)}_\mathit{e\sigma}
    \cdot
    \mathbf{g}^\mathit{(f)}_\mathit{e\zeta}( \mathbf{\Tilde{x}}^\mathit{(f)}_\mathit{e\sigma})
    =
    \delta_\mathit{\zeta\sigma}$.
The form of $ \mathbf{g}^\mathit{(f)}_\mathit{e\zeta} $ determines a particular FR scheme. 
Further details on the correction function can be found in \cite{vincent2011}.

Combining the above, Equation \ref{eq:transq} can be expressed as
\begin{equation}\label{eq:FR-qfinal}
    \Tilde{\mathbf{q}}^\mathit{(u)}_\mathit{e\zeta n\alpha}
    =
    \Big[ \hat{\Tilde{\mathbf{n}}}^\mathit{(f)}_\mathit{e\sigma}
    \Tilde{\nabla}\cdot\Tilde{\mathbf{g}}^\mathit{(f)}_\mathit{e\sigma}
        (\Tilde{\mathbf{x}})
        \big\{
        \mathfrak{C}_\alpha u^{(f)}_{\mathit{e\sigma n\alpha}}
        - u^{(f)}_{\mathit{e\sigma n\alpha}}
        \big\}
        + \Tilde{\mathbf{u}}^\mathit{(u)}_\mathit{e\nu n\alpha}
            \cdot \Tilde{\nabla}\ell^\mathit{(u)}_\mathit{e\nu}
                (\Tilde{\mathbf{x}})
    \Big]_{\Tilde{\mathbf{x}}=\Tilde{\mathbf{x}}^\mathit{(u)}_\mathit{e\zeta}}.
\end{equation}
This equation simply computes the transformed gradient by using the analytic derivation of the basis polynomial $\ell_e$ together with the solution at solution points, and adds the contribution of the jump between the discontinuous solution at the flux points and the common solution at flux points.

The physical gradients at solution points and at flux points can be computed using approaches of Kopriva \cite{KoprivaDavidA1998ASMS} and Sun et al. \cite{SunWangLiu2007} as
\begin{equation}\label{eq:FR-gradcoru}
\mathbf{q}^\mathit{(u)}_\mathit{e\zeta\!n\!\alpha}
=
\mathbf{J}_{en}^{-T}(\mathbf{\Tilde{x}}_{e\sigma}^{(u)})
 \Tilde{\mathbf{q}}^\mathit{(u)}_\mathit{e\zeta\!n\!\alpha},
\end{equation}
\begin{equation}\label{eq:FR-gradcoru_fpts}
\mathbf{q}^\mathit{(f)}_\mathit{e\zeta\!n\!\alpha}
=
\ell_{e\zeta}^{(u)}(\mathbf{\Tilde{x}}_{e\sigma}^{(f)}) \Tilde{\mathbf{q}}^\mathit{(u)}_\mathit{e\zeta n\alpha}.
\end{equation}

The auxiliary equation is now solved and transformed flux can be obtained as
\begin{equation}\label{eq:FR-tdisf}
\Tilde{\mathbf{f}}_{e\sigma n\alpha}^{(u)}
=
J_{e\sigma n}^{(u)} \mathbf{J}_{e\sigma n}^{-1(u)} 
  \mathbf{f}_\alpha(u_{e\sigma n}^{(u)}, \mathbf{q}_{e\sigma n}^{(u)}),
\end{equation}
where $J_{e\sigma n}^{(u)} = \mathsf{det}\big(\mathbf{J}_{en}^{-T}(\mathbf{\Tilde{x}}_{e\sigma}^{(u)})\big)$.
Now a similar procedure used in Equation \ref{eq:FR-disu} can be applied here to obtain the normal transformed flux at flux points as
\begin{equation}
f^\mathit{(f_\perp)}_\mathit{e\sigma n\alpha}
=
\ell^{(u)}_\mathit{e\zeta}
      (\Tilde{\mathbf{x}}^{(f)}_\mathit{e\sigma})
      \hat{\Tilde{\mathbf{n}}}^\mathit{(f)}_\mathit{e\sigma}
      \cdot \Tilde{\mathbf{f}}_{e\zeta n\alpha}^{(u)}.
\end{equation}

With the discontinuous solution and the physical gradient at the flux points, the next step is to obtain the common normal flux for each flux point pair in the domain. 
The inviscid component of the flux is computed using a Riemann solver \cite{ToroRiemann, Rusanov}.
The viscous component of the flux is obtained by using the Local Discontinuous Galerkin (LDG) method as described in \cite{hesthaven2007nodal} and it is defined as
\begin{equation}
\mathfrak{F}^v(\mathbf{f}^v_L, \mathbf{f}^v_R, \hat{\mathbf{n}}_L) = 
\hat{\mathbf{n}}_L\cdot 
\big(
  \big( \sfrac{1}{2} + \beta \big) \mathbf{f}_L^v 
+ \big( \sfrac{1}{2} - \beta \big) \mathbf{f}_R^v.
\big) + \tau (u_L - u_R),
\end{equation}
where $\tau$ is a penalty parameter, $\mathbf{f}_L^v = \mathbf{f}^v(u_L, \mathbf{q}_L)$, $\mathbf{f}_R^v = \mathbf{f}^v(u_R, \mathbf{q}_R)$, and $\beta$ controls the upwinding or downwinding.

$\mathfrak{F}_\alpha$ is the operator for the common normal flux at the interface, and $f^\mathit{(f_\perp)}$ denotes the normal flux at a flux point.
Note that interfaces between elements conform and we have $\hat{\mathbf{n}}_{\sigma n}^{(f)} = -\hat{\mathbf{n}}_{\sigma' n'}^{(f)}$. 
Due to the conservation property of the numerical scheme, we always have 
$\mathfrak{F}_\alpha f^\mathit{(f_\perp)}_\mathit{\sigma\!n\!\alpha}
=
-\mathfrak{F}_\alpha f^\mathit{(f_\perp)}_{\widetilde{\mathit{\sigma\!n\!\alpha}}}$, and the full form of the common normal flux operator can be given as, 

\begin{equation}\label{eq:FR-comm_flux}
    \mathfrak{F}_\alpha f^\mathit{(f_\perp)}_\mathit{e\sigma n\alpha}
    =
    -\mathfrak{F}_\alpha f^\mathit{(f_\perp)}_{\widetilde{\mathit{e\sigma n\alpha}}}
    =
    \mathfrak{F}_\alpha
    (u^\mathit{(f)}_\mathit{e\sigma n},
    u^\mathit{(f)}_{\widetilde{\mathit{e\sigma n}}},
    \mathbf{q}^\mathit{(f)}_\mathit{e\sigma n},
    \mathbf{q}^\mathit{(f)}_{\widetilde{\mathit{e\sigma n}}},
    \hat{\mathbf{n}}^\mathit{(f)}_\mathit{e\sigma n}).
\end{equation}

Next, we need to obtain the flux at solution points $\Tilde{\mathbf{f}}^\mathit{(u)}_\mathit{e\sigma n\alpha}$ from the solution state at solution points $\mathbf{u}^\mathit{(u)}_\mathit{e\sigma n\alpha}$ using Equations \ref{eq:fluxe} and \ref{eq:fluxv}.
The flux at solution points will then be used to obtain the normal flux at flux points $f^\mathit{(f_\perp)}_\mathit{e\sigma n\alpha}$.
The difference between the flux and the common normal flux at flux points is used in the correction procedure analogous to the Equation \ref{eq:FR-qfinal}.
Finally, assembling all of the steps defined above, we can define an expression for the divergence of the continuous flux as,
\begin{equation}\label{eq:FRfinal}
    (\Tilde{\nabla}\cdot\Tilde{\mathbf{f}})^\mathit{(u)}_\mathit{e\zeta n\alpha}
    =
    \Big[
    \Tilde{\nabla}\cdot\Tilde{\mathbf{g}}^\mathit{(f)}_\mathit{e\sigma}
        (\Tilde{\mathbf{x}})
        \big\{
        \mathfrak{F}_\alpha f^\mathit{(f_\perp)}_\mathit{e\sigma n\alpha}
        - f^\mathit{(f_\perp)}_\mathit{e\sigma n\alpha}
        \big\}
        + \Tilde{\mathbf{f}}^\mathit{(u)}_\mathit{e\nu n\alpha}
            \cdot \Tilde{\nabla}\ell^\mathit{(u)}_\mathit{e\nu}
                (\Tilde{\mathbf{x}})
    \Big]_{\Tilde{\mathbf{x}}=\Tilde{\mathbf{x}}^\mathit{(u)}_\mathit{e\zeta}},
\end{equation}
which results in a semi-discretised form of the governing system.

\subsubsection{Anti-aliasing formulation}
FR uses collocation based projections to obtain the transformed flux at the solution points.
This approach is efficient but may result in aliasing errors where the energy from unresolved modes gets erroneously transferred into resolved modes and consequently this results in stability problems as well as accuracy issues.
One way to overcome aliasing errors is using $L^2$ based projections with suitable quadrature rules \cite{Park17}.

Two types of anti-aliasing are considered here, flux anti-aliasing and surface-flux anti-aliasing.
Flux anti-aliasing uses fluxes at quadrature points to perform an L2 projection of flux and then samples this projection at the original solution points. 
Surface-flux anti-aliasing applies the flux correction operation in a similar way. First, fluxes at flux points are used to obtain the fluxes at the quadrature points at the interfaces. Then, common flux and common solution are evaluated at the quadrature points at the interfaces, and finally common flux and solution are sampled at the original flux points. 
These procedures require extra operators and changes in some of the existing operators. 
First, flux anti-aliasing requires two new matrix multiplication operations. 
One of them operates on the solution and the other operates on the gradients at the solution points.
Next, surface-flux anti-aliasing requires changes in many of the existing operators in the FR algorithm. 

First, for flux anti-aliasing, interpolation from solution points to quadrature points can be expressed as
\begin{equation}\label{eq:FR-uqpts}
    u^{(q)}_\mathit{e\sigma n\alpha}
    =
    u^{(u)}_\mathit{e\zeta n\alpha}\ell^{(u)}_\mathit{e\zeta}
      (\Tilde{\mathbf{x}}^{(q)}_\mathit{e\sigma})
\end{equation}
\begin{equation}\label{eq:FR-gradcoru_qpts}
    \mathbf{q}^{(q)}_\mathit{e\sigma n\alpha}
    =
    \mathbf{q}^{(u)}_\mathit{e\zeta n\alpha}\ell^{(u)}_\mathit{e\zeta}
      (\Tilde{\mathbf{x}}^{(q)}_\mathit{e\sigma})
\end{equation}

Then the solution and gradients at the quadrature points are used to evaluate the flux at quadrature points, replacing Equation \ref{eq:FR-tdisf} thus
\begin{equation}\label{eq:FR-tdisfq}
\Tilde{\mathbf{f}}_{e\sigma n\alpha}^{(q)}
=
J_{e\sigma n}^{(u)} \mathbf{J}_{e\sigma n}^{-1(q)} 
  \mathbf{f}_\alpha(u_{e\sigma n}^{(q)}, \mathbf{q}_{e\sigma n}^{(q)}).
\end{equation}
Then, in order to sample the fluxes at the original solution points the following operation can be used
\begin{equation}\label{eq:FR-tdivtpcorfq}
\Tilde{\mathbf{f}}_{e\zeta n\alpha}^{(u)}
=
\psi_{e\nu}(\Tilde{\mathbf{x}}^{(u)}_\mathit{e\zeta})\omega^{(q)}_\mathit{e\sigma}
\psi_{e\nu}(\Tilde{\mathbf{x}}^{(q)}_\mathit{e\sigma})
\Tilde{\mathbf{f}}_{e\sigma n\alpha}^{(u)}.
\end{equation}

Next, a similar strategy can be used for the surface-flux anti-aliasing.
Surface-flux anti-aliasing evaluates the solution and flux at the quadrature points, obtains the common flux and common solution, and samples these back at the original flux points. 
However, instead of bringing up more operations to carry out the interpolations, surface-flux anti-aliasing can be implemented simply by replacing flux points by quadrature points at the interfaces. 
This will result in changes in all the operations that include data at the flux points yet the formulation will be the same as presented in the previous section.

\subsection{Formulation in Terms of Kernels}
The operations required to implement the FR algorithm described in Section \ref{sec:FR} can be cast in terms of individual kernels.
These can be grouped into two main types, matrix-matrix multiplication and point-wise kernels.
In PyFR, various third party libraries such as LIBXSMM \cite{xsmm} and GiMMiK \cite{gimmik} are made available for the matrix-matrix multiplication kernels.
Point-wise kernels are implemented through PyFR-Mako, a bespoke templating language derived from Mako, which is then rendered into low-level platform-specific code.

In order to formulate the FR algorithm in terms of kernels, we define the various constant operator matrices and point-wise kernels, alongside various storage matrices for state variables and some intermediate data.

First, we define the \textit{disu} kernel, which corresponds to Equation \ref{eq:FR-disu}.
This kernel evaluates the solution at flux points using the solution at solution points, and can be formulated as a matrix multiplication. 
The constant operator matrix of element type $e$ for the \textit{disu} kernel, $\mathbf{M}_e^0$, is defined as
\begin{equation}
    (\mathbf{M}_e^0)_\mathit{\sigma\zeta} 
    =
    \ell^{(u)}_\mathit{e\zeta}(\Tilde{\mathbf{x}}^\mathit{(f)}_\mathit{e\sigma}), \qquad \dim\  \mathbf{M}_e^0 = N^\mathit{(f)}_e \times N^\mathit{(u)}_e,
\end{equation}
the array $\mathbf{U}_e^\mathit{(u)}$ for storing the solution state at solution points is defined as
\begin{equation}
    (\mathbf{U}_e^\mathit{(u)})_\mathit{\zeta(n\alpha)}
    =
    u^\mathit{(u)}_\mathit{e \zeta n\alpha}, \qquad
    \dim \mathbf{U}_e^\mathit{(u)} 
       = N_e^\mathit{(u)} \times N_V  |\mathbf{\Omega}_e|.
\end{equation}
Therefore, the \textit{disu} kernel can be cast as a matrix multiplication as,
\begin{equation}
    \mathbf{U}_e^\mathit{(f)}
    =
    \mathbf{M}_e^0
    \mathbf{U}_e^\mathit{(u)},
\end{equation}
where $\mathbf{U}_e^\mathit{(f)}$ is the storage for the solution state at flux points which has dimension $\dim \mathbf{U}_e^\mathit{(f)} = N_e^\mathit{(f)} \times N_V |\mathbf{\Omega}_e|$.

The next five kernels are dedicated to obtaining the gradients of the solution which is a requirement for calculating the flux and common flux.
The first operation related to the gradients is obtaining the common solution at the flux points which is defined in Equation \ref{eq:FR-conu}.
The \textit{con\_u} kernel is defined to obtain the common solution as
\begin{equation}
    \mathbf{C}_e^\mathit{(f)} = \textit{con\_u}(\mathbf{U}_e^\mathit{(f)}, \mathbf{n}_e^\mathit{(f)}),
\end{equation}
which is a point-wise kernel with considerable indirect memory access to main memory.
The \textit{con\_u} kernel uses up-winding or down-winding to determine the common solution at the element interfaces, and writes the solution into a sub-region of the $\mathbf{F}_e^\mathit{(f)}$ array in order to save memory usage.
For clarity we will refer to this sub-region of the $\mathbf{F}_e^\mathit{(f)}$ matrix as $\mathbf{C}_e^\mathit{(f)}$ after its data is replaced by the \textit{con\_u} kernel.
$\mathbf{n}_e^\mathit{(f)}$ refers to the outward pointing normal vector at the flux points.

Afterwards, the corrections have to be applied in order to obtain the gradients.
The procedure to apply corrections is described in Equation \ref{eq:FR-qfinal}.
Therefore the following two constant matrices are defined thus
\begin{alignat}{3}
    (\mathbf{M}_e^4)_\mathit{\zeta\sigma} 
    &=
    \big[ \Tilde{\nabla} \ell^\mathit{(u)}_\mathit{e\zeta} (\Tilde{\mathbf{x}}) 
    \big]^T_{\Tilde{\mathbf{x}}=\Tilde{\mathbf{x}}^\mathit{(u)}_\mathit{e\sigma}} , \qquad &&\dim \mathbf{M}_e^4 &&= N_e^\mathit{(u)} \times N_D N_e^\mathit{(u)},\\
    (\mathbf{M}_e^6)_\mathit{\zeta\sigma} 
    &=
    \big[\Tilde{\hat{\mathbf{n}}}^{(f)}_\mathit{e\zeta} \cdot \Tilde{\nabla} \cdot \mathbf{g}^\mathit{(f)}_\mathit{e\zeta} (\Tilde{\mathbf{x}}) 
    \big]_{\Tilde{\mathbf{x}}=\Tilde{\mathbf{x}}^\mathit{(u)}_\mathit{e\sigma}} , \quad &&\dim \mathbf{M}_e^6 &&= N_e^\mathit{(u)} \times N_e^\mathit{(f)}.
\end{alignat}
Using the matrices defined above, Equation \ref{eq:FR-qfinal} can be written in matrix multiplication form as
\begin{equation}\label{eq:Qsubs}
\begin{split}
    \Tilde{\mathbf{Q}}_e^\mathit{(u)} &= \mathbf{M}_e^6 \{\Tilde{\mathbf{C}}_e^\mathit{(f)}-\Tilde{\mathbf{U}}_e^\mathit{(f)}\}
               + \mathbf{M}_e^4 \Tilde{\mathbf{U}}_e^\mathit{(u)} \\
               &= \mathbf{M}_e^6 \{\Tilde{\mathbf{C}}_e^\mathit{(f)}-\mathbf{M}_e^0 \Tilde{\mathbf{U}}_e^\mathit{(u)}\}
               + \mathbf{M}_e^4 \Tilde{\mathbf{U}}_e^\mathit{(u)} \\
               &= \mathbf{M}_e^6 \Tilde{\mathbf{C}}_e^\mathit{(f)} - \{\mathbf{M}_e^4 - \mathbf{M}_e^6\mathbf{M}_e^0\} \Tilde{\mathbf{U}}_e^\mathit{(u)},
\end{split}
\end{equation}
where the $\Tilde{\mathbf{Q}}_e^\mathit{(u)}$ array is the storage for the gradients at solution points and has a size $\dim\  \Tilde{\mathbf{Q}}_e^\mathit{(u)} = N_D N_e^\mathit{(u)} \times N_V |\mathbf{\Omega}_e|$.
The size of $\Tilde{\mathbf{Q}}_e^\mathit{(u)}$ is $N_D$ times bigger compared to that of $\Tilde{\mathbf{U}}_e^\mathit{(u)}$ as there are $N_D$ many components of gradient for each state variable.
As there are two arrays and two constant matrix multiplications involved in this operation, it can be cast as two consecutive matrix multiplication kernels.
Corresponding to the first and second terms in the equation, \textit{tgradcoru} and \textit{tgradpcoru} kernels are defined respectively.
The \textit{tgradpcoru} is called first and the output is stored at $\Tilde{\mathbf{Q}}_e^\mathit{(u)}$ thus
\begin{equation}
\Tilde{\mathbf{Q}}_e^\mathit{(u)} = \{\mathbf{M}_e^4 - \mathbf{M}_e^6\mathbf{M}_e^0\} \Tilde{\mathbf{U}}_e^\mathit{(u)}.
\end{equation}
Next, the \textit{tgradcoru} kernel is called and the result is added to $\Tilde{\mathbf{Q}}_e^\mathit{(u)}$ as
\begin{equation}
\Tilde{\mathbf{Q}}_e^\mathit{(u)} = \Tilde{\mathbf{Q}}_e^\mathit{(u)} +
\mathbf{M}_e^6 \Tilde{\mathbf{C}}_e^\mathit{(f)}.
\end{equation}

Then, a point-wise kernel, \textit{gradcoru} is defined in order to obtain the gradients of the solution at solution points in physical space as formulated in Equation \ref{eq:FR-gradcoru} thus
\begin{equation}
    \mathbf{Q}_e^\mathit{(u)} = \textit{gradcoru}(\mathbf{J}_e^{(u)},\Tilde{\mathbf{Q}}_e^\mathit{(u)}).
\end{equation}

Afterwards, the \textit{gradcoru\_fpts} kernel is defined in order to evaluate the physical gradients at the flux points corresponding to Equation \ref{eq:FR-gradcoru_fpts}.
This is also a matrix multiplication kernel and the associated constant operator matrix is $\mathbf{M}_e^5$, which is defined as
\begin{equation}
    (\mathbf{M}_e^5)_\mathit{\sigma\zeta} 
    =
    \text{diag}(\mathbf{M}_e^0, ..., \mathbf{M}_e^0), \qquad \text{dim}\  \mathbf{M}_e^5 = N_D N_e^\mathit{(f)} \times N_D N_e^\mathit{(u)}.
\end{equation}
Therefore, \textit{gradcoru\_fpts} kernel can be defined as
\begin{equation}
\mathbf{Q}_e^\mathit{(f)} = \mathbf{M}_e^5 \mathbf{Q}_e^\mathit{(u)},
\end{equation}
where $\mathbf{Q}_e^\mathit{(f)}$ is the storage array for the gradients at flux points with the dimension $\text{dim}\ \mathbf{Q}_e^\mathit{(f)} = N_D N_e^\mathit{(f)} \times N_V |\mathbf{\Omega}_e|$.

Next, with he solution and its gradient available at the element interfaces, it is now possible to obtain the common interface flux at the element interfaces as in Equation \ref{eq:FR-comm_flux}. 
For this purpose, the \textit{comm\_flux} kernel is defined thus
\begin{equation}
    \mathbf{D}_e^\mathit{(f)} = \textit{comm\_flux}(\mathbf{U}_e^\mathit{(f)}, \mathbf{Q}_e^\mathit{(f)}, \mathbf{n}_e^\mathit{(f)}),
\end{equation}
which is a point-wise kernel with a significant indirect memory access requirement.
The \textit{comm\_flux} kernel calls a Riemann flux function on pairs of flux points reading $\mathbf{U}_e^\mathit{(f)}$ and $\mathbf{F}_e^\mathit{(f)}$, then returning $\mathbf{D}_e^\mathit{(f)}$.
In practice it overwrites the data over $\mathbf{U}_e^\mathit{(f)}$.
However for clarity, we will refer to the $\mathbf{U}_e^\mathit{(f)}$ matrix as $\mathbf{D}_e^\mathit{(f)}$ after its data is replaced by the \textit{comm\_flux} kernel.

Then, the next step is obtaining the transformed flux at the solution points as in Equation \ref{eq:FR-tdisf}.
A point-wise kernel, \textit{tdisf}, is defined to evaluate the flux at solution points thus
\begin{equation}
    \Tilde{\mathbf{F}}_e^\mathit{(u)}
    =
    \textit{tdisf} (
    \mathbf{U}_e^\mathit{(u)}, \mathbf{Q}_e^\mathit{(u)}, \mathbf{x}_e^\mathit{(u)}
    ).
\end{equation}
This kernel takes the solution $\mathbf{U}_e^\mathit{(u)}$, gradient at the solution points $\mathbf{Q}_e^\mathit{(u)}$, and the element coordinates $\mathbf{x}_e^\mathit{(u)}$, and returns the transformed flux $\Tilde{\mathbf{F}}_e^\mathit{(u)}$, where $\Tilde{\mathbf{F}}_e^\mathit{(u)}$ has a size $\text{dim}\ \Tilde{\mathbf{F}}^\mathit{(u)} = N_D N_e^\mathit{(u)} \times N_V |\mathbf{\Omega}_e|$.
Because the gradients at the solution points are not needed any more, the $\mathbf{Q}_e^\mathit{(u)}$ array is overwritten by $\mathbf{F}_e^\mathit{(u)}$ and the array is referred as $\mathbf{F}_e^\mathit{(u)}$ from this point onwards.

In order to rewrite the final semi-discrete form in Equation \ref{eq:FRfinal} in matrix multiplication format, the following constant operator matrices are defined,
\begin{alignat}{3}
    (\mathbf{M}_e^1)_\mathit{\zeta\sigma} 
    &=
    \big[ \Tilde{\nabla} \ell^\mathit{(u)}_\mathit{e\zeta} (\Tilde{\mathbf{x}}) 
    \big]^T_{\Tilde{\mathbf{x}}=\Tilde{\mathbf{x}}^\mathit{(u)}_\mathit{e\sigma}} , \qquad &&\dim \mathbf{M}_e^1 &&= N_e^\mathit{(u)} \times N_D N_e^\mathit{(u)}\\
    (\mathbf{M}_e^2)_\mathit{\zeta\sigma} 
    &=
    \big[ \ell^\mathit{(u)}_\mathit{e\zeta} (\Tilde{\mathbf{x}}^\mathit{(f)}_\mathit{e\sigma}) \Hat{\Tilde{\mathbf{n}}}^\mathit{(f)}_\mathit{e\sigma}
    \big]^T , \qquad &&\dim \mathbf{M}_e^2 &&= N_e^\mathit{(f)} \times N_D N_e^\mathit{(u)}\\
    (\mathbf{M}_e^3)_\mathit{\zeta\sigma} 
    &=
    \big[ \Tilde{\nabla} \cdot \mathbf{g}^\mathit{(f)}_\mathit{e\zeta} (\Tilde{\mathbf{x}}) 
    \big]_{\Tilde{\mathbf{x}}=\Tilde{\mathbf{x}}^\mathit{(u)}_\mathit{e\sigma}} , \quad &&\dim \mathbf{M}_e^3 &&= N_e^\mathit{(u)} \times N_e^\mathit{(f)}.
\end{alignat}

Then Equation \ref{eq:FRfinal} can be rewritten using the matrices defined above as
\begin{equation}\label{eq:subs}
\begin{split}
    \Tilde{\mathbf{R}}_e^\mathit{(u)} &= \mathbf{M}_e^3 \{\Tilde{\mathbf{D}}_e^\mathit{(f)}-\mathbf{M}_e^2\Tilde{\mathbf{F}}_e^\mathit{(u)}\}
               + \mathbf{M}_e^1 \Tilde{\mathbf{F}}_e^\mathit{(u)} \\
               &= \mathbf{M}_e^3 \Tilde{\mathbf{D}}_e^\mathit{(f)}+ \{\mathbf{M}_e^1 - \mathbf{M}_e^3\mathbf{M}_e^2\}\Tilde{\mathbf{F}}_e^\mathit{(u)},
\end{split}
\end{equation}
where the $\Tilde{\mathbf{R}}_e^\mathit{(u)}$ array has a size $\dim \Tilde{\mathbf{R}}_e^\mathit{(u)} = N_e^\mathit{(u)} \times N_V |\mathbf{\Omega}_e|$, equivalent to the size of the solution array $\mathbf{U}_e^\mathit{(u)}$.
The kernel implementation of Equation \ref{eq:FRfinal} consists of two matrix multiplication kernels that correspond to the two terms in the right hand side of the equation, \textit{tdivtconf} and \textit{tdivpcorf} respectively.
The \textit{tdivpcorf} kernel is called first and the data in the $\Tilde{\mathbf{R}}_e^\mathit{(u)}$ array is overwritten by the result as
\begin{equation}
    \Tilde{\mathbf{R}}_e^\mathit{(u)}
    =
    \{\mathbf{M}_e^1 - \mathbf{M}_e^3\mathbf{M}_e^2\}
    \Tilde{\mathbf{F}}_e^\mathit{(u)}.
\end{equation}
Next, the \textit{tdivtconf} kernel is executed and the result is added to $\Tilde{\mathbf{R}}_e^\mathit{(u)}$ thus
\begin{equation}
    \Tilde{\mathbf{R}}_e^\mathit{(u)}
    =
    \Tilde{\mathbf{R}}_e^\mathit{(u)}
    +
    \mathbf{M}_e^3 \Tilde{\mathbf{D}}_e^\mathit{(f)}.
\end{equation}

Finally, the \textit{negdivconf} kernel is defined in order to obtain the divergence of flux in the physical space thus
\begin{equation}
    \mathbf{R}_e^\mathit{(u)} = \textit{negdivconf}(\det \mathbf{J}_e^{(u)},\Tilde{\mathbf{R}}_e^\mathit{(u)}).
\end{equation}

\subsubsection{Anti-aliasing Kernels}
\paragraph{Flux anti-aliasing}
Flux anti-aliasing requires two additional kernels which evaluate the solution state and gradients at quadrature points within each element, such that the \textit{tdisf} kernel can then evaluate these fluxes at these same quadrature points. 
For this purpose \textit{uqpts} and \textit{gradcoru\_qpts} kernels are defined which interpolate the solution and gradients at solution points to quadrature points.
They are both matrix multiplication kernels corresponding to Equations \ref{eq:FR-uqpts} and \ref{eq:FR-gradcoru_qpts} respectively.
The associated operator matrix $\mathbf{M}_e^7$ for the \textit{uqpts} kernel is defined as
\begin{equation}
    (\mathbf{M}_e^7)_\mathit{\sigma\zeta} 
    =
    \ell_\mathit{e\zeta}^{(u)}(\Tilde{\mathbf{x}}_\mathit{e\sigma}^\mathit{(q)}), \qquad \text{dim}\  \mathbf{M}_e^7 = N_e^\mathit{(u)} \times N_e^\mathit{(q)},
\end{equation}
and the \textit{uqpts} kernel is defined as
\begin{equation}
    \mathbf{U}_e^\mathit{(q)}
    =
    \mathbf{M}_e^7
    \mathbf{U}_e^\mathit{(u)}.
\end{equation}
The operator matrix $\mathbf{M}_e^{10}$ for \textit{gradcoru\_qpts} is given as
\begin{equation}
    (\mathbf{M}_e^{10})_\mathit{\sigma\zeta} 
    =
    \text{diag}(\mathbf{M}_e^7, ..., \mathbf{M}_e^7), \qquad \text{dim}\  \mathbf{M}_e^{10} = N_D N_e^\mathit{(u)} \times N_D N_e^\mathit{(q)},
\end{equation}
and the \textit{gradcoru\_qpts} kernel can be expressed as
\begin{equation}
    \mathbf{Q}_e^\mathit{(q)}
    =
    \mathbf{M}_e^{10}
    \mathbf{Q}_e^\mathit{(u)}.
\end{equation}

When flux anti-aliasing is turned on, the \textit{tdisf} kernel evaluates the flux at quadrature points using the solution and gradients at quadrature points thus it can be redefined as
\begin{equation}
    \Tilde{\mathbf{F}}_e^\mathit{(q)}
    =
    \textit{tdisf} (
    \mathbf{U}_e^\mathit{(q)}, \mathbf{Q}_e^\mathit{(q)}, \mathbf{x}_e
    ),
\end{equation}
which obtains the modal coefficients in the solution space.
Then, the operator matrix $\mathbf{M}_e^8$ is defined to convert these modal coefficients into nodal values at the solution points.
\begin{equation}
    (\mathbf{M}_e^8)_\mathit{\sigma\zeta} 
    =
    \psi_{e\nu}(\Tilde{\mathbf{x}}^{(u)}_\mathit{e\zeta})\omega^{(q)}_\mathit{e\sigma}
    \psi_{e\nu}(\Tilde{\mathbf{x}}^{(q)}_\mathit{e\sigma}),
 \qquad \text{dim}\  \mathbf{M}_e^8 = N_e^\mathit{(q)} \times N_e^\mathit{(u)}.
\end{equation}
$\mathbf{M}_e^9$ is the block diagonal form of the $\mathbf{M}_e^8$ matrix and it is defined as
\begin{equation}
    (\mathbf{M}_e^9)_\mathit{\sigma\zeta} 
    =
    \text{diag}(\mathbf{M}_e^8, ..., \mathbf{M}_e^8), \qquad \dim\  \mathbf{M}_e^9 = N_D N_e^\mathit{(q)} \times N_D N_e^\mathit{(u)}.
\end{equation}
This operation is carried out by the \textit{tdivtpcorf} kernel with a small update therefore when flux anti-aliasing is in use \textit{tdivtpcorf} kernel becomes
\begin{equation}
    \Tilde{\mathbf{R}}_e^\mathit{(u)}
    =
    \{\mathbf{M}_e^1 - \mathbf{M}_e^3\mathbf{M}_e^2\}\mathbf{M}_e^9
    \Tilde{\mathbf{F}}_e^\mathit{(q)}.
\end{equation}

\begin{figure}[h!]
  \centering
\begin{tikzpicture}

\def\yloc{0}
\def\name{negdivconf}
\def\cntr{0}
\node[draw, fill=red!50] (\name) at (\cntr,\yloc) {\name};
\node (\name_l) at (\cntr-2,\yloc) {$\mathbf{R}^{(u)}$};
\node (\name_r) at (\cntr+2,\yloc) {$\mathbf{R}^{(u)}$}; 
\node[draw,dotted,fit=(\name) (\name_l) (\name_r)] {};
\path[->, very thick] (\name_l) edge (\name);
\path[->, very thick] (\name) edge (\name_r);

\def\yloc{1.5}
\def\name{tdivtconf}
\def\cntr{0}
\node[draw, fill=blue!75!green!40] (\name) at (\cntr,\yloc) {\name};
\node (tdivtconf_ll) at (\cntr-3,\yloc) {$\mathbf{U}^{(f)}$,};
\node (tdivtconf_l) at (\cntr-2,\yloc) {$\mathbf{R}^{(u)}$};
\node (tdivtconf_r) at (\cntr+2,\yloc) {$\mathbf{R}^{(u)}$}; 
\node[draw,dotted,fit=(\name) (\name_l) (\name_r) (\name_ll)] {};
\path[->, very thick] (\name_l) edge (\name);
\path[->, very thick] (\name) edge (\name_r);

\def\yloc{3}
\def\name{tdivpcorf}
\def\cntr{4}
\node[draw, fill=blue!75!green!40] (\name) at (\cntr,\yloc) {\name};
\node (\name_l) at (\cntr-2,\yloc) {$\mathbf{F}^{(q)}$};
\node (\name_r) at (\cntr+2,\yloc) {$\mathbf{R}^{(u)}$}; 
\node[draw,dotted,fit=(\name) (\name_l) (\name_r)] {};
\path[->, very thick] (\name_l) edge (\name);
\path[->, very thick] (\name) edge (\name_r);

\def\yloc{4.5}
\def\name{tdisf}
\def\cntr{5}
z\node[draw, fill=red!50] (\name) at (\cntr,\yloc) {\name};
\node (\name_ll) at (\cntr-3,\yloc) {$\mathbf{F}^{(q)},$};
\node (\name_l) at (\cntr-2,\yloc) {$\mathbf{U}^{(q)}$};
\node (\name_r) at (\cntr+2,\yloc) {$\mathbf{F}^{(q)}$};
\node[draw,dotted,fit=(\name) (\name_ll) (\name_l) (\name_r)] {};
\path[->, very thick] (\name_l) edge (\name);
\path[->, very thick] (\name) edge (\name_r);

\def\yloc{6.5}
\def\name{qpts}
\def\cntr{5}
\node[draw, fill=blue!75!green!40] (\name) at (\cntr,\yloc) {\name};
\node (\name_ll) at (\cntr-2.5,\yloc) {$\mathbf{F}^{(u)}$,};
\node (\name_l) at (\cntr-1.5,\yloc) {$\mathbf{U}^{(u)}$};
\node (\name_r) at (\cntr+1.5,\yloc) {$\mathbf{F}^{(q)}$,};
\node (\name_rr) at (\cntr+2.5,\yloc) {$\mathbf{U}^{(q)}$};
\begin{scope}[on background layer]
\node[draw,dotted,fit=(\name) (\name_ll) (\name_l) (\name_r) (\name_rr)] {};
\end{scope}
\path[->, very thick] (\name_l) edge (\name);
\path[->, very thick] (\name) edge (\name_r);

\def\yloc{4.5}
\def\name{comm_flux}
\def\cntr{-3}
\node[draw, fill=pink] (\name) at (\cntr,\yloc) {comm\_flux};
\node (\name_ll) at (\cntr-3,\yloc) {$\mathbf{U}^{(f)},$};
\node (\name_l) at (\cntr-2,\yloc) {$\mathbf{F}^{(f)}$};
\node (\name_r) at (\cntr+2,\yloc) {$\mathbf{U}^{(f)}$}; 
\node[draw,dotted,fit=(\name) (\name_ll) (\name_l) (\name_r)] {};
\path[->, very thick] (\name_l) edge (\name);
\path[->, very thick] (\name) edge (\name_r);

\def\yloc{13}
\def\name{disu}
\def\cntr{-3}
\node[draw, fill=blue!75!green!40] (\name) at (\cntr,\yloc) {\name};
\node (\name_l) at (\cntr-2,\yloc) {$\mathbf{U}^{(u)}$};
\node (\name_r) at (\cntr+2,\yloc) {$\mathbf{U}^{(f)}$}; 
\node[draw,dotted,fit=(\name) (\name_l) (\name_r)] {};
\path[->, very thick] (\name_l) edge (\name);
\path[->, very thick] (\name) edge (\name_r);

\def\yloc{6.5}
\def\name{gradcoru_fpts}
\def\cntr{-2}
\node[draw, fill=blue!75!green!40] (\name) at (\cntr,\yloc) {gradcoru\_fpts};
\node (\name_l) at (\cntr-2,\yloc) {$\mathbf{F}^{(u)}$};
\node (\name_r) at (\cntr+2,\yloc) {$\mathbf{F}^{(f)}$}; 
\node[draw,dotted,fit=(\name) (\name_l) (\name_r)] {};
\path[->, very thick] (\name_l) edge (\name);
\path[->, very thick] (\name) edge (\name_r);

\def\yloc{8}
\def\name{gradcoru}
\def\cntr{0}
\node[draw, fill=red!50] (\name) at (\cntr,\yloc) {\name};
\node (\name_l) at (\cntr-2,\yloc) {$\mathbf{F}^{(u)}$};
\node (\name_r) at (\cntr+2,\yloc) {$\mathbf{F}^{(u)}$}; 
\node[draw,dotted,fit=(\name) (\name_l) (\name_r)] {};
\path[->, very thick] (\name_l) edge (\name);
\path[->, very thick] (\name) edge (\name_r);

\def\yloc{11}
\def\name{tgradpcoru}
\def\cntr{3}
\node[draw, fill=blue!75!green!40] (\name) at (\cntr,\yloc) {\name};
\node (\name_l) at (\cntr-2,\yloc) {$\mathbf{U}^{(u)}$};
\node (\name_r) at (\cntr+2,\yloc) {$\mathbf{F}^{(u)}$}; 
\node[draw,dotted,fit=(\name) (\name_l) (\name_r)] {};
\path[->, very thick] (\name_l) edge (\name);
\path[->, very thick] (\name) edge (\name_r);

\def\yloc{9.5}
\def\name{tgradcoru}
\def\cntr{0}
\node[draw, fill=blue!75!green!40] (\name) at (\cntr,\yloc) {\name};
\node (\name_ll) at (\cntr-3,\yloc) {$\mathbf{U}^{(f)}$};
\node (\name_l) at (\cntr-2,\yloc) {$\mathbf{F}^{(u)}$};
\node (\name_r) at (\cntr+2,\yloc) {$\mathbf{F}^{(u)}$}; 
\node[draw,dotted,fit=(\name) (\name_l) (\name_r) (\name_ll)] {};
\path[->, very thick] (\name_l) edge (\name);
\path[->, very thick] (\name) edge (\name_r);

\def\yloc{11}
\def\name{con_u}
\def\cntr{-3}
\node[draw, fill=pink] (\name) at (\cntr,\yloc) {con\_u};
\node (\name_l) at (\cntr-2,\yloc) {$\mathbf{U}^{(f)}$};
\node (\name_r) at (\cntr+2,\yloc) {$\mathbf{U}^{(f)}$}; 
\node[draw,dotted,fit=(\name) (\name_l) (\name_r)] {};
\path[->, very thick] (\name_l) edge (\name);
\path[->, very thick] (\name) edge (\name_r);

\draw[->, very thick, looseness=1] (disu_r.south west) .. controls(-6.25,12.5) .. (comm_flux_ll);
\draw[->, very thick, looseness=0.75] (disu_r) to [out=-90,in=90] (con_u_l);
\draw[->, very thick, looseness=0.75] (comm_flux_r) to [out=-90,in=90] (tdivtconf_ll);
\draw[->, very thick, looseness=0.75] (tdivtconf_r) to [out=-90,in=90] (negdivconf_l);

\draw[->, very thick, looseness=0.75] (tdisf_r) to [out=-90,in=90] (tdivpcorf_l);
\draw[->, very thick, looseness=0.5] (tdivpcorf_r) to [out=-90,in=90] (tdivtconf_l);

\draw[->, very thick, looseness=0.75] (tgradcoru_r) to [out=-90,in=90] (gradcoru_l);
\draw[->, very thick, looseness=0.5] (tgradpcoru_r) to [out=-90,in=90] (tgradcoru_l);

\draw[->, very thick, looseness=0.75] (con_u_r) to [out=-90,in=90] (tgradcoru_ll);

\draw[->, very thick, looseness=0.5] (gradcoru_r) to [out=-90,in=90] (gradcoru_fpts_l);


\draw[->, very thick, looseness=0.75] (gradcoru_fpts_r) to [out=-90,in=90] (comm_flux_l.north);

\node (Uin) at (0,14) {$\mathbf{U}^{(u)}$};
\node (Uout) at (3,-1) {$\partial \mathbf{U}^{(u)} / \partial t$};
\draw[->, very thick] (Uin) to [out=180,in=90, looseness=0.5] (disu_l);
\draw[->, very thick] (Uin) to [out=-45,in=90, looseness=0.5] (tgradpcoru_l);
\draw[->, very thick] (negdivconf_r) to [out=0,in=90] (Uout);

\draw[->, very thick, looseness=2] (Uin) to [out=0,in=45] (qpts_l.north);
\draw[->, very thick, looseness=1] (gradcoru_r) to [out=-50,in=90] (qpts_ll.north);
\draw[->, very thick, looseness=0.5] (qpts_r.-135) to [out=-120,in=90] (tdisf_ll.90);
\draw[->, very thick, looseness=0.45] (qpts_rr.-120) to [out=-90,in=90] (tdisf_l.90);

\node[draw, below left,inner sep=2, fill=blue!75!green!40,minimum size=2cm,minimum size=12,minimum width=30] (a) at (-4.0, -1.0) {};
\node[draw, below left,inner sep=2, fill=red!50,minimum size=2cm,,minimum size=12,minimum width=30] (b) at (a.south east) {};
\node[draw, below left,inner sep=2, fill=pink ,minimum size=12,minimum width=30] (c) at (b.south east) {};

\node [below right] (x) at (-4.0, -1.0) {\footnotesize Matrix Multiplication Kernel};
\node [below right] (y) at (x.south west) {\footnotesize Point-wise Kernel};
\node [below right] (z) at (y.south west) {\footnotesize Point-wise Kernel with Indirect Memory Access};

\end{tikzpicture}

  \caption{Kernel prerequisites map for Navier-Stokes Solver including optional Flux AA kernels.}
  \label{fig:NSAAkernels}
\end{figure}
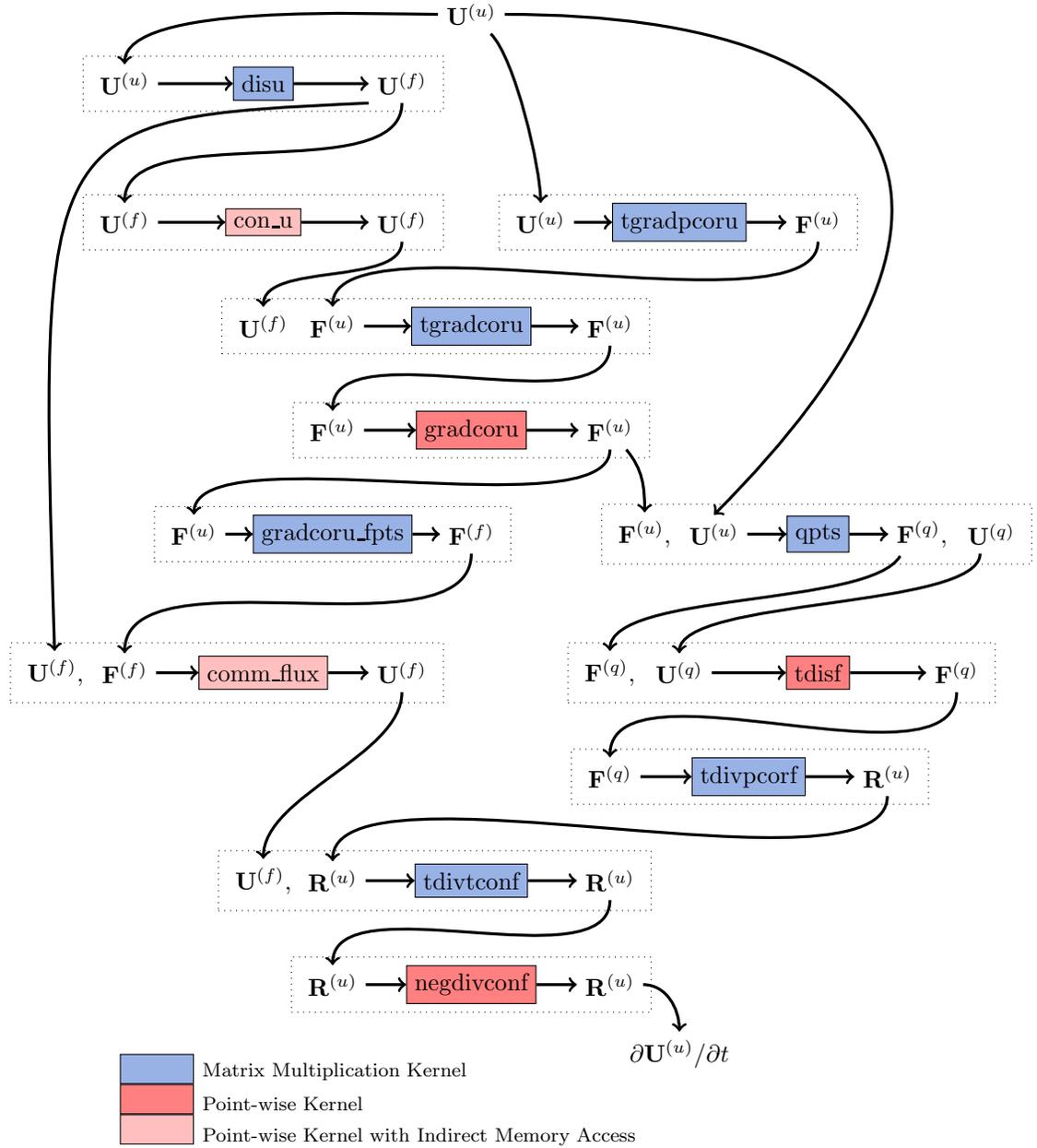

\paragraph{Surface-flux anti-aliasing}
Surface-flux anti-aliasing on the other hand changes many of the constant operator matrices that act on flux points. 
Because the formulation is valid for any flux point set, turning on surface-flux anti-aliasing does not alter any of the kernel or the formulation, however it changes the data movement requirements. 
A schematic that includes all the Navier-Stokes kernels and the additional flux anti-aliasing kernels is given in Figure \ref{fig:NSAAkernels}.

\section{Kernel Execution Order and Data Movement Requirements}\label{sec:kex}
As a baseline for comparison, data movement and bandwidth requirements are first analysed for PyFR version v1.11.0, which has no cache blocking support.
Similar to analysis carried out for the Euler solver in \cite{Akkurt}, time integration related operations are not included in data movement requirements and speedups therefore will only reflect the performance of evaluating the RHS.
In reality, a time integrator such as RK4 will need to be used to move forward in time which requires additional data movement that consumes bandwidth and it will slightly reduce the performance per RHS call.

\begin{figure}[p!]
    \centering
\scalebox{0.92}{
\begin{tikzpicture}

\def\yloc{0}
\def\name{negdivconf}
\node[draw, fill=red!50] (\name) at (0,\yloc) {\name};
\node (\name_l) at (-2,\yloc) {$\mathbf{R}^{(u)}$};
\node (\name_r) at (2,\yloc) {$\mathbf{R}^{(u)}$}; 
\node[draw,dotted,fit=(\name) (\name_l) (\name_r)] {};
\path[->, very thick] (\name_l) edge (\name);
\path[->, very thick] (\name) edge (\name_r);

\draw[->, very thick] (\name_r) to [out=0,in=-180] (4,\yloc);
\draw[->, very thick] (-5,\yloc) to [out=0,in=-180] (\name_l);

\node (one) at (-5.75,\yloc) {13};
\draw[gray,dashed] (-4,0.75) -- (4,\yloc+0.75);

\def\yloc{1.5}
\def\name{tdivtconf}
\node[draw, fill=blue!75!green!40] (\name) at (0,\yloc) {\name};
\node (\name_l) at (-2,\yloc) {$\mathbf{U}^{(f)}$};
\node (\name_r) at (2,\yloc) {$\mathbf{R}^{(u)}$}; 
\node[draw,dotted,fit=(\name) (\name_l) (\name_r)] {};
\path[->, very thick] (\name_l) edge (\name);
\path[->, very thick] (\name) edge (\name_r);

\draw[->, very thick] (\name_r) to [out=0,in=-180] (4,\yloc);
\draw[->, very thick] (-5,\yloc) to [out=0,in=-180] (\name_l);

\node (one) at (-5.75,\yloc) {12};
\draw[gray,dashed] (-4,\yloc+0.75) -- (4,\yloc+0.75);

\def\yloc{3}
\node[draw, fill=pink] (comm_flux) at (0,\yloc) {comm\_flux};
\node (comm_flux_ll) at (-3,\yloc) {$\mathbf{U}^{(f)}$,};
\node (comm_flux_l) at (-2,\yloc) {$\mathbf{F}^{(f)}$};
\node (comm_flux_r) at (2,\yloc) {$\mathbf{U}^{(f)}$}; 
\node[draw,dotted,fit=(comm_flux)(comm_flux_ll) (comm_flux_l) (comm_flux_r)] {};
\path[->, very thick] (comm_flux_l) edge (comm_flux);
\path[->, very thick] (comm_flux) edge (comm_flux_r);

\draw[->, very thick] (comm_flux_r) to [out=0,in=-180] (4,\yloc);
\draw[->, very thick] (-5,\yloc) to [out=0,in=-180] (comm_flux_ll);

\node (one) at (-5.75,\yloc) {11};
\draw[gray,dashed] (-4,\yloc+0.75) -- (4,\yloc+0.75);

\def\yloc{4.5}
\def\name{tdivpcorf}
\node[draw, fill=blue!75!green!40] (\name) at (0,\yloc) {\name};
\node (\name_l) at (-2,\yloc) {$\mathbf{F}^{(q)}$};
\node (\name_r) at (2,\yloc) {$\mathbf{R}^{(u)}$}; 
\node[draw,dotted,fit=(\name) (\name_l) (\name_r)] {};
\path[->, very thick] (\name_l) edge (\name);
\path[->, very thick] (\name) edge (\name_r);

\draw[->, very thick] (\name_r) to [out=0,in=-180] (4,\yloc);
\draw[->, very thick] (-5,\yloc) to [out=0,in=-180] (\name_l);

\node (one) at (-5.75,\yloc) {10};
\draw[gray,dashed] (-4,\yloc+0.75) -- (4,\yloc+0.75);

\def\yloc{6}
\def\name{tdisf}
\node[draw, fill=red!50] (\name) at (0,\yloc) {\name};
\node (\name_ll) at (-3,\yloc) {$\mathbf{U}^{(q)}$,};
\node (\name_l) at (-2,\yloc) {$\mathbf{F}^{(q)}$};
\node (\name_r) at (2,\yloc) {$\mathbf{F}^{(q)}$}; 
\node[draw,dotted,fit=(\name) (\name_ll) (\name_l) (\name_r)] {};
\path[->, very thick] (\name_l) edge (\name);
\path[->, very thick] (\name) edge (\name_r);

\draw[->, very thick] (\name_r) to [out=0,in=-180] (4,\yloc);
\draw[->, very thick] (-5,\yloc) to [out=0,in=-180] (\name_ll);

\node (one) at (-5.75,\yloc) {9};
\draw[gray,dashed] (-4,\yloc+ 0.75) -- (4,\yloc+0.75);

\def\yloc{7.5}
\node[draw, fill=blue!75!green!40] (gradcoru_q) at (0,\yloc) {gradcoru\_qpts};
\node (gradcoru_q_l) at (-2,\yloc) {$\mathbf{F}^{(u)}$};
\node (gradcoru_q_r) at (2,\yloc) {$\mathbf{F}^{(q)}$}; 
\node[draw,dotted,fit=(gradcoru_q) (gradcoru_q_l) (gradcoru_q_r)] {};
\path[->, very thick] (gradcoru_q_l) edge (gradcoru_q);
\path[->, very thick] (gradcoru_q) edge (gradcoru_q_r);

\draw[->, very thick] (gradcoru_q_r) to [out=0,in=-180] (4,\yloc);
\draw[->, very thick] (-5,\yloc) to [out=0,in=-180] (gradcoru_q_l);

\node (one) at (-5.75,\yloc) {7};
\draw[gray,dashed] (-4,\yloc+0.75) -- (4,\yloc+0.75);

\def\yloc{9}
\node[draw, fill=blue!75!green!40] (qptsu) at (0,\yloc) {qptsu};
\node (qptsu_l) at (-2,\yloc) {$\mathbf{U}^{(u)}$};
\node (qptsu_r) at (2,\yloc) {$\mathbf{U}^{(q)}$}; 
\node[draw,dotted,fit=(qptsu) (qptsu_l) (qptsu_r)] {};
\path[->, very thick] (qptsu_l) edge (qptsu);
\path[->, very thick] (qptsu) edge (qptsu_r);

\draw[->, very thick] (qptsu_r) to [out=0,in=-180] (4,\yloc);
\draw[->, very thick] (-5,\yloc) to [out=0,in=-180] (qptsu_l);

\node (one) at (-5.75,\yloc) {8};
\draw[gray,dashed] (-4,\yloc+0.75) -- (4,\yloc+0.75);

\def\yloc{10.5}
\node[draw, fill=blue!75!green!40] (gradcoru_fpts) at (0,\yloc) {gradcoru\_fpts};
\node (gradcoru_fpts_l) at (-2,\yloc) {$\mathbf{F}^{(u)}$};
\node (gradcoru_fpts_r) at (2,\yloc) {$\mathbf{F}^{(f)}$}; 
\node[draw,dotted,fit=(gradcoru_fpts) (gradcoru_fpts_l) (gradcoru_fpts_r)] {};
\path[->, very thick] (gradcoru_fpts_l) edge (gradcoru_fpts);
\path[->, very thick] (gradcoru_fpts) edge (gradcoru_fpts_r);

\draw[->, very thick] (gradcoru_fpts_r) to [out=0,in=-180] (4,\yloc);
\draw[->, very thick] (-5,\yloc) to [out=0,in=-180] (gradcoru_fpts_l);

\node (one) at (-5.75,\yloc) {6};
\draw[gray,dashed] (-4,\yloc+0.75) -- (4,\yloc+0.75);

\def\yloc{12}
\def\name{gradcoru}
\node[draw, fill=red!50] (\name) at (0,\yloc) {\name};
\node (\name_l) at (-2,\yloc) {$\mathbf{F}^{(u)}$};
\node (\name_r) at (2,\yloc) {$\mathbf{F}^{(u)}$}; 
\node[draw,dotted,fit=(\name) (\name_l) (\name_r)] {};
\path[->, very thick] (\name_l) edge (\name);
\path[->, very thick] (\name) edge (\name_r);

\draw[->, very thick] (\name_r) to [out=0,in=-180] (4,\yloc);
\draw[->, very thick] (-5,\yloc) to [out=0,in=-180] (\name_l);

\node (one) at (-5.75,\yloc) {5};
\draw[gray,dashed] (-4,\yloc+0.75) -- (4,\yloc+0.75);

\def\yloc{13.5}
\def\name{tgradcoru}
\node[draw, fill=blue!75!green!40] (\name) at (0,\yloc) {\name};
\node (\name_l) at (-2,\yloc) {$\mathbf{U}^{(f)}$};
\node (\name_r) at (2,\yloc) {$\mathbf{F}^{(u)}$}; 
\node[draw,dotted,fit=(\name) (\name_l) (\name_r)] {};
\path[->, very thick] (\name_l) edge (\name);
\path[->, very thick] (\name) edge (\name_r);

\draw[->, very thick] (\name_r) to [out=0,in=-180] (4,\yloc);
\draw[->, very thick] (-5,\yloc) to [out=0,in=-180] (\name_l);

\node (one) at (-5.75,\yloc) {4};
\draw[gray,dashed] (-4,\yloc+0.75) -- (4,\yloc+0.75);

\def\yloc{15}
\def\name{tgradpcoru}
\node[draw, fill=blue!75!green!40] (\name) at (0,\yloc) {\name};
\node (\name_l) at (-2,\yloc) {$\mathbf{U}^{(u)}$};
\node (\name_r) at (2,\yloc) {$\mathbf{F}^{(u)}$}; 
\node[draw,dotted,fit=(\name) (\name_l) (\name_r)] {};
\path[->, very thick] (\name_l) edge (\name);
\path[->, very thick] (\name) edge (\name_r);

\draw[->, very thick] (\name_r) to [out=0,in=-180] (4,\yloc);
\draw[->, very thick] (-5,\yloc) to [out=0,in=-180] (\name_l);

\node (one) at (-5.75,\yloc) {3};
\draw[gray,dashed] (-4,\yloc+0.75) -- (4,\yloc+0.75);

\def\yloc{16.5}
\def\name{conu}
\node[draw, fill=pink] (con_u) at (0,\yloc) {con\_u};
\node (con_u_l) at (-2,\yloc) {$\mathbf{U}^{(f)}$};
\node (con_u_r) at (2,\yloc) {$\mathbf{U}^{(f)}$}; 
\node[draw,dotted,fit=(con_u) (con_u_l) (con_u_r)] {};
\path[->, very thick] (con_u_l) edge (con_u);
\path[->, very thick] (con_u) edge (con_u_r);

\draw[->, very thick] (con_u_r) to [out=0,in=-180] (4,\yloc);
\draw[->, very thick] (-5,\yloc) to [out=0,in=-180] (con_u_l);

\node (one) at (-5.75,\yloc) {2};
\draw[gray,dashed] (-4,\yloc+0.75) -- (4,\yloc+0.75);

\def\yloc{18}
\def\name{disu}
\node[draw, fill=blue!75!green!40] (\name) at (0,\yloc) {\name};
\node (\name_l) at (-2,\yloc) {$\mathbf{U}^{(u)}$};
\node (\name_r) at (2,\yloc) {$\mathbf{U}^{(f)}$}; 
\node[draw,dotted,fit=(\name) (\name_l) (\name_r)] {};
\path[->, very thick] (\name_l) edge (\name);
\path[->, very thick] (\name) edge (\name_r);

\draw[->, very thick] (\name_r) to [out=0,in=-180] (4,\yloc);
\draw[->, very thick] (-5,\yloc) to [out=0,in=-180] (\name_l);

\node (one) at (-5.75,\yloc) {1};

\definecolor{mygreen}{RGB}{160,213,104}
\draw[mygreen, very thick] (-4.25,-0.5) -- (-4.25,18.5);
\draw[mygreen, very thick] (3.25,-0.5) -- (3.25,18.5);

\end{tikzpicture}}
    \caption{Current kernel execution order. Arrows going over green line indicate main memory access.}
    \label{fig:NSkernelorder}
\end{figure}
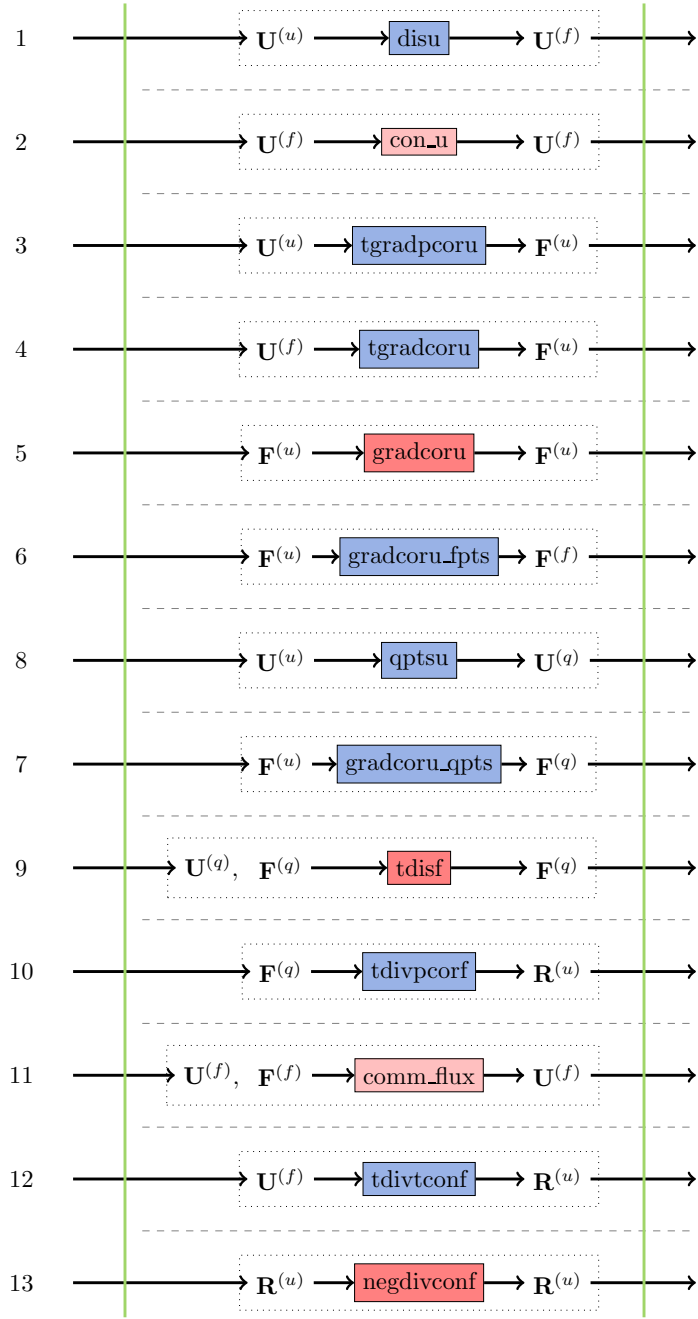

Kernel execution order for the Navier-Stokes solver in PyFR version v1.11.0 is demonstrated in Figure \ref{fig:NSkernelorder} including the flux anti-aliasing kernels.
Without cache blocking support, each kernel is executed for the entire input and output arrays and therefore the data movement requirement is directly related to the size of input and output arrays.
Data movement requirements of each kernel are tabulated in Table \ref{table:NSBW} for a hexahedral mesh, normalized per element, at polynomial orders $p=3$ and $p=4$, double precision, and including flux and surface-flux anti-aliasing on/off.
Non-temporal stores are used whenever possible for the matrix multiplication kernels through the LIBXSMM library \cite{xsmm} to reduce the data movement requirements for the qualifying kernels.
Total data movement requirements for polynomial orders $p=3$ and $p=4$ for anti-aliasing on and off is summarised in Table \ref{table:NSBWsum}.

\begin{table}[h!]
\caption{Total data movement requirement for PyFR version v1.11.0.}\label{table:NSBWsum}
\centering
\scalebox{0.8}{
\begin{tabular}{rr|r|r|r|r|r|r|r}
 \multicolumn{9}{r}{Data Movement Req. Per RHS Per Element [KiB/RHS/element]} \\ \cline{2-9}
 & \multicolumn{4}{c|}{AA off} & \multicolumn{4}{|c}{Full AA} \\ \cline{2-9}
 & \multicolumn{2}{c|}{Hexa} & \multicolumn{2}{|c|}{Prism} & \multicolumn{2}{|c|}{Hexa} & \multicolumn{2}{|c}{Prism} \\ \cline{2-9}
  & $p=3$ & $p=4$ & $p=3$ & $p=4$ & $p=3$ & $p=4$ & $p=3$ & $p=4$       \\ \hline 
         \multicolumn{1}{r|}{Total Bandwidth}      & 143.63 & 276.35 & 102.44 & 163.88 & 224.41 & 381.12 & 144.80 & 238.05 \\  
\end{tabular}
}
\end{table}

\section{Tensor Product Factorisation of Anti-aliasing Operators}\label{sec:AA}

Turning on anti-aliasing in PyFR brings in additional operators and changes in the existing kernels.
A straightforward implementation of these new operators and changes in the existing kernels results in dense matrix operators.
However, it is well known in the literature that for tensor product elements these operators can be cast as multiple sparse factors as opposed to a single dense matrix multiplication. 
Converting a single dense operator into a chain of sparse operations reduces the FLOP requirement significantly, but a direct implementation can make the code overly bandwidth bound and the performance gains would be limited. 
However for tensor product elements such as hexahedra and prisms, it is possible to factor the dense operator into sparse components.
In this section, we examine ways to factor the new dense operators associated with flux anti-aliasing and the existing operators that are converted into a dense form due to surface-flux anti-aliasing.
A breakdown of all the affected kernels due to flux and surface-flux anti-aliasing is shown in Table \ref{table:operatorms}.

\begin{table}[!h]
\caption{Changes in operator matrices due to anti-aliasing. Blue colour indicates a change due to flux anti-aliasing, and red colour indicates a change due to surface-flux anti-aliasing.}\label{table:operatorms}
\centering
\scalebox{0.85}{
\begin{tabular}{l||c||c|c|c}
         \multicolumn{2}{r}{} & \multicolumn{3}{c}{Changes in the kernels due to AA} \\ \cline{3-5}
Kernel    & Operator & {\color[HTML]{3531FF}flux AA}                            & {\color[HTML]{FE0000}surf-flux AA}                   & full AA                            \\ \hline
\textit{disu}           & $\mathbf{M}^0$       &                                    & {\color[HTML]{FE0000} $\mathbf{M}^0$}      & {\color[HTML]{FE0000} $\mathbf{M}^0$}          \\ 
\textit{con\_u}         & Pointwise       &                                    &                                &                                    \\ 
\textit{tgradpcoru}     & $\mathbf{M}^4-\mathbf{M}^6\mathbf{M}^0$  &                                    & $\mathbf{M}^4-${\color[HTML]{FE0000} $\mathbf{M}^6\mathbf{M}^0$} & $\mathbf{M}^4$-{\color[HTML]{FE0000} $\mathbf{M}^6\mathbf{M}^0$}     \\ 
\textit{tgradcoru}      & $\mathbf{M}^6$       &                                    & {\color[HTML]{FE0000} $\mathbf{M}^6$}      & {\color[HTML]{FE0000} $\mathbf{M}^6$}          \\ 
\textit{gradcoru}       & Pointwise       &                                    &                                &                                    \\ 
\textit{gradcoru\_fpts} & $I_3 \bigotimes \mathbf{M}^0$  &                                    & $I_3 \bigotimes \textcolor{red}{\mathbf{M}^0}$      & $I_3 \bigotimes \textcolor{red}{\mathbf{M}^0}$  \\ 
\textit{comm\_flux}     & Pointwise       &                                    &                              &                                    \\ 
\textit{uqpts}          &  $\mathbf{M}^7$ & {\color[HTML]{3531FF} $\mathbf{M}^7$}          &                                & {\color[HTML]{3531FF} $\mathbf{M}^7$}          \\ 
\textit{gradcoru\_qpts} & $I_3 \bigotimes \mathbf{M}^7$ & $I_3 \bigotimes \textcolor{blue}{\mathbf{M}^7}$     &                                & $I_3 \bigotimes \textcolor{blue}{\mathbf{M}^7}$     \\ 
\textit{tdisf}          & Pointwise       &                                    &                                &                                    \\ 
\textit{tdivtpcorf}     & $\mathbf{M}^1-\mathbf{M}^3\mathbf{M}^2$  & ($\mathbf{M}^1-\mathbf{M}^3\mathbf{M}^2$){\color[HTML]{3531FF} $\mathbf{M}^9$} & $\mathbf{M}^1-${\color[HTML]{FE0000} $\mathbf{M}^3\mathbf{M}^2$}                    & ($\mathbf{M}^1-${\color[HTML]{FE0000} $\mathbf{M}^3\mathbf{M}^2$}){\color[HTML]{3531FF} $\mathbf{M}^9$} \\ 
\textit{tdivtconf}      & $\mathbf{M}^3$       &                                    & {\color[HTML]{FE0000} $\mathbf{M}^3$}      & {\color[HTML]{FE0000} $\mathbf{M}^3$}          \\ 
\textit{negdivconf}     & Pointwise       &                                    &                                &                                   
\end{tabular}
}
\end{table}

First, flux anti-aliasing brings in two new operators, $\mathbf{M}^7$ and $\mathbf{M}^9$.
$\mathbf{M}^7$ operator interpolates data in solution points to quadrature points.
If implemented directly, it is a single dense operator.
However, for a hexahedral element this process can be factored into three sparse components.
The process is explained in detail in \cite{Warburton19}.
The sparse components obtained after factorisation correspond to interpolations in $x, y,$ and $z$ directions.
Sampling the nodal coefficients at solution points using the modal representation at quadrature points can also be factored into three sparse components.
$\mathbf{M}^9$ operator applies the reverse operation for the gradients at solution points, so its a block diagonal matrix that applies the reverse interpolation for three components of the gradient at each solution point.

The interpolations between solution points and quadrature points are the main building blocks for decomposing the rest of the operators that are associated with the surface-flux anti-aliasing.
The only extra operation involved with surface-flux anti-aliasing is the interpolation to flux points at the interfaces.
The $\mathbf{M}^0$ operator without anti-aliasing that interpolates data from solution points to flux points will be used together with the $\mathbf{M}^7$ or $\mathbf{M}^9$ operator to form the sparse factors of the $\mathbf{M}^0$ operator with surface-flux anti-aliasing.

\subsection{Extension to Prism Elements}
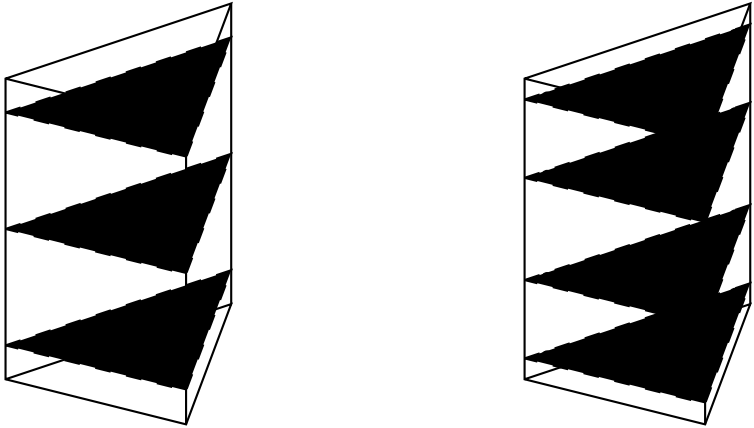
\begin{figure}[h!]
    \centering
\scalebox{2}{
\begin{tikzpicture}[x = {(0.75cm,0.25cm)}, y={(0.6cm,-0.15cm)}, z={(0cm,1cm)},]
\draw [color=orange, mark=*, mark size=1] plot coordinates{(-0.81684757298044, 0.63369514596088, -0.7745966692414834)};
\draw [color=orange, mark=*, mark size=1] plot coordinates{(0.63369514596088, -0.81684757298044, -0.7745966692414834)};
\draw [color=orange, mark=*, mark size=1] plot coordinates{(-0.81684757298044, -0.81684757298044, -0.7745966692414834)};
\draw [color=orange, mark=*, mark size=1] plot  coordinates{(-0.108103018168072, -0.783793963663856, -0.7745966692414834)};
\draw [color=orange, mark=*, mark size=1] plot coordinates{(-0.783793963663856, -0.108103018168072, -0.7745966692414834)};
\draw [color=orange, mark=*, mark size=1] plot coordinates{(-0.108103018168072, -0.108103018168072, -0.7745966692414834)};
\draw [color=orange, mark=*, mark size=1] plot coordinates{(-0.81684757298044, 0.63369514596088, 0.0)};
\draw [color=orange, mark=*, mark size=1] plot coordinates{(0.63369514596088, -0.81684757298044, 0.0)};
\draw [color=orange, mark=*, mark size=1] plot coordinates{(-0.81684757298044, -0.81684757298044, 0.0)};
\draw [color=orange, mark=*, mark size=1] plot coordinates{(-0.108103018168072, -0.783793963663856, 0.0)};
\draw [color=orange, mark=*, mark size=1] plot coordinates{(-0.783793963663856, -0.108103018168072, 0.0)};
\draw [color=orange, mark=*, mark size=1] plot coordinates{(-0.108103018168072, -0.108103018168072, 0.0)};
\draw [color=orange, mark=*, mark size=1] plot coordinates{(-0.81684757298044, 0.63369514596088, 0.7745966692414834)};
\draw [color=orange, mark=*, mark size=1] plot coordinates{(0.63369514596088, -0.81684757298044, 0.7745966692414834)};
\draw [color=orange, mark=*, mark size=1] plot coordinates{(-0.81684757298044, -0.81684757298044, 0.7745966692414834)};
\draw [color=orange, mark=*, mark size=1] plot coordinates{(-0.108103018168072, -0.783793963663856, 0.7745966692414834)};
\draw [color=orange, mark=*, mark size=1] plot coordinates{(-0.783793963663856, -0.108103018168072, 0.7745966692414834)};
\draw [color=orange, mark=*, mark size=1] plot coordinates{(-0.108103018168072, -0.108103018168072, 0.7745966692414834)};
\path (-1,-1,-1) coordinate (A) (1,-1,-1) coordinate (B) (-1,1,-1) coordinate (C); 
\path (-1,-1,1) coordinate (D) (1,-1,1) coordinate (E) (-1,1,1) coordinate (F);
\draw (A)--(B)--(C)--(A)--(D)--(E)--(F)--(D);
\draw (B)--(E);
\draw (C)--(F);
\foreach \z in {-0.7745966692414834,0,0.7745966692414834}{
\draw [dashed, black, fill=black, fill opacity=0.075] (-1,-1,\z)--(1,-1,\z)--(-1,1,\z)--(-1,-1,\z);
}
\end{tikzpicture}
\qquad
\hspace{1cm}
\begin{tikzpicture}[x = {(0.75cm,0.25cm)}, y={(0.6cm,-0.15cm)}, z={(0cm,1cm)},]
\draw plot [mark=*, mark size=1] coordinates{(-0.333333333333333, -0.333333333333333, -0.8611363115940526)};
\draw plot [mark=*, mark size=1] coordinates{(-0.888871894660414, 0.777743789320828, -0.8611363115940526)};
\draw plot [mark=*, mark size=1] coordinates{(0.777743789320828, -0.888871894660414, -0.8611363115940526)};
\draw plot [mark=*, mark size=1] coordinates{(-0.888871894660414, -0.888871894660414, -0.8611363115940526)};
\draw plot [mark=*, mark size=1] coordinates{(0.268421495491446, -0.408932576528214, -0.8611363115940526)};
\draw plot [mark=*, mark size=1] coordinates{(-0.859488918963232, -0.408932576528214, -0.8611363115940526)};
\draw plot [mark=*, mark size=1] coordinates{(-0.408932576528214, -0.859488918963232, -0.8611363115940526)};
\draw plot [mark=*, mark size=1] coordinates{(0.268421495491446, -0.859488918963232, -0.8611363115940526)};
\draw plot [mark=*, mark size=1] coordinates{(-0.859488918963232, 0.268421495491446, -0.8611363115940526)};
\draw plot [mark=*, mark size=1] coordinates{(-0.408932576528214, 0.268421495491446, -0.8611363115940526)};
\draw plot [mark=*, mark size=1] coordinates{(-0.333333333333333, -0.333333333333333, -0.33998104358485626)};
\draw plot [mark=*, mark size=1] coordinates{(-0.888871894660414, 0.777743789320828, -0.33998104358485626)};
\draw plot [mark=*, mark size=1] coordinates{(0.777743789320828, -0.888871894660414, -0.33998104358485626)};
\draw plot [mark=*, mark size=1] coordinates{(-0.888871894660414, -0.888871894660414, -0.33998104358485626)};
\draw plot [mark=*, mark size=1] coordinates{(0.268421495491446, -0.408932576528214, -0.33998104358485626)};
\draw plot [mark=*, mark size=1] coordinates{(-0.859488918963232, -0.408932576528214, -0.33998104358485626)};
\draw plot [mark=*, mark size=1] coordinates{(-0.408932576528214, -0.859488918963232, -0.33998104358485626)};
\draw plot [mark=*, mark size=1] coordinates{(0.268421495491446, -0.859488918963232, -0.33998104358485626)};
\draw plot [mark=*, mark size=1] coordinates{(-0.859488918963232, 0.268421495491446, -0.33998104358485626)};
\draw plot [mark=*, mark size=1] coordinates{(-0.408932576528214, 0.268421495491446, -0.33998104358485626)};
\draw [color=orange] plot [mark=*, mark size=1] coordinates{(-0.333333333333333, -0.333333333333333, 0.33998104358485626)};
\draw plot [mark=*, mark size=1] coordinates{(-0.888871894660414, 0.777743789320828, 0.33998104358485626)};
\draw plot [mark=*, mark size=1] coordinates{(0.777743789320828, -0.888871894660414, 0.33998104358485626)};
\draw plot [mark=*, mark size=1] coordinates{(-0.888871894660414, -0.888871894660414, 0.33998104358485626)};
\draw plot [mark=*, mark size=1] coordinates{(0.268421495491446, -0.408932576528214, 0.33998104358485626)};
\draw plot [mark=*, mark size=1] coordinates{(-0.859488918963232, -0.408932576528214, 0.33998104358485626)};
\draw plot [mark=*, mark size=1] coordinates{(-0.408932576528214, -0.859488918963232, 0.33998104358485626)};
\draw plot [mark=*, mark size=1] coordinates{(0.268421495491446, -0.859488918963232, 0.33998104358485626)};
\draw plot [mark=*, mark size=1] coordinates{(-0.859488918963232, 0.268421495491446, 0.33998104358485626)};
\draw plot [mark=*, mark size=1] coordinates{(-0.408932576528214, 0.268421495491446, 0.33998104358485626)};
\draw plot [mark=*, mark size=1] coordinates{(-0.333333333333333, -0.333333333333333, 0.8611363115940526)};
\draw plot [mark=*, mark size=1] coordinates{(-0.888871894660414, 0.777743789320828, 0.8611363115940526)};
\draw plot [mark=*, mark size=1] coordinates{(0.777743789320828, -0.888871894660414, 0.8611363115940526)};
\draw plot [mark=*, mark size=1] coordinates{(-0.888871894660414, -0.888871894660414, 0.8611363115940526)};
\draw plot [mark=*, mark size=1] coordinates{(0.268421495491446, -0.408932576528214, 0.8611363115940526)};
\draw plot [mark=*, mark size=1] coordinates{(-0.859488918963232, -0.408932576528214, 0.8611363115940526)};
\draw plot [mark=*, mark size=1] coordinates{(-0.408932576528214, -0.859488918963232, 0.8611363115940526)};
\draw plot [mark=*, mark size=1] coordinates{(0.268421495491446, -0.859488918963232, 0.8611363115940526)};
\draw plot [mark=*, mark size=1] coordinates{(-0.859488918963232, 0.268421495491446, 0.8611363115940526)};
\draw plot [mark=*, mark size=1] coordinates{(-0.408932576528214, 0.268421495491446, 0.8611363115940526)};
\path (-1,-1,-1) coordinate (A) (1,-1,-1) coordinate (B) (-1,1,-1) coordinate (C); 
\path (-1,-1,1) coordinate (D) (1,-1,1) coordinate (E) (-1,1,1) coordinate (F);
\draw (A)--(B)--(C)--(A)--(D)--(E)--(F)--(D);
\draw (B)--(E);
\draw (C)--(F);
\foreach \z in {-0.8611363115940526,-0.33998104358485626,0.33998104358485626,0.8611363115940526}{
\draw [dashed, black, fill=black, fill opacity=0.075] (-1,-1,\z)--(1,-1,\z)--(-1,1,\z)--(-1,-1,\z);
}
\end{tikzpicture}
}
    \caption{Solution points for prism elements with polynomial orders $p=2$ and $p=3$.
Point distribution is based on William-Shun at the triangular cross sections and based on Gauss-Legendre along the z-direction.
A direct interpolation between $p=2$ and $p=3$ prisms solution points results in a dense operator matrix as any point in $p=3$ depend on all the points in $p=2$ as indicated with orange points.}
    \label{fig:prism2toprims3}
\end{figure}

The decomposition into sparse factors is apparent for hexahedral elements, however, there is no generic way of obtaining sparse factors for tetrahedral or pyramid elements.
Prisms on the other hand are slightly different.
A direct approach for obtaining the modal coefficients at quadrature points by using the nodal coefficients at solution points in a prism would require a dense operator matrix as shown in Figure \ref{fig:prism2toprims3}.

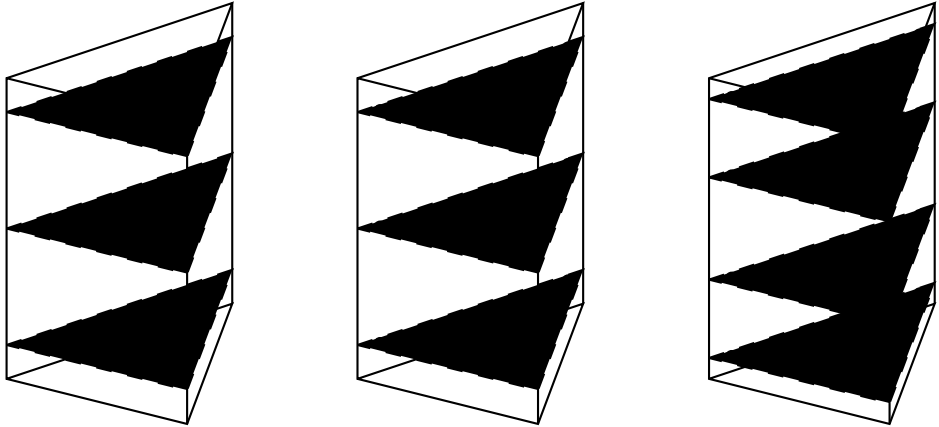
\begin{figure}[h!]
    \centering
\scalebox{2}{
\begin{tikzpicture}[x = {(0.75cm,0.25cm)}, y={(0.6cm,-0.15cm)}, z={(0cm,1cm)},]
\draw plot [mark=*, mark size=1] coordinates{(-0.81684757298044, 0.63369514596088, -0.7745966692414834)};
\draw plot [mark=*, mark size=1] coordinates{(0.63369514596088, -0.81684757298044, -0.7745966692414834)};
\draw plot [mark=*, mark size=1] coordinates{(-0.81684757298044, -0.81684757298044, -0.7745966692414834)};
\draw plot [mark=*, mark size=1] coordinates{(-0.108103018168072, -0.783793963663856, -0.7745966692414834)};
\draw plot [mark=*, mark size=1] coordinates{(-0.783793963663856, -0.108103018168072, -0.7745966692414834)};
\draw plot [mark=*, mark size=1] coordinates{(-0.108103018168072, -0.108103018168072, -0.7745966692414834)};
\draw plot [mark=*, mark size=1] coordinates{(-0.81684757298044, 0.63369514596088, 0.0)};
\draw plot [mark=*, mark size=1] coordinates{(0.63369514596088, -0.81684757298044, 0.0)};
\draw plot [mark=*, mark size=1] coordinates{(-0.81684757298044, -0.81684757298044, 0.0)};
\draw plot [mark=*, mark size=1] coordinates{(-0.108103018168072, -0.783793963663856, 0.0)};
\draw plot [mark=*, mark size=1] coordinates{(-0.783793963663856, -0.108103018168072, 0.0)};
\draw plot [mark=*, mark size=1] coordinates{(-0.108103018168072, -0.108103018168072, 0.0)};
\draw [color=orange] plot [mark=*, mark size=1] coordinates{(-0.81684757298044, 0.63369514596088, 0.7745966692414834)};
\draw [color=orange] plot [mark=*, mark size=1] coordinates{(0.63369514596088, -0.81684757298044, 0.7745966692414834)};
\draw [color=orange] plot [mark=*, mark size=1] coordinates{(-0.81684757298044, -0.81684757298044, 0.7745966692414834)};
\draw [color=orange] plot [mark=*, mark size=1] coordinates{(-0.108103018168072, -0.783793963663856, 0.7745966692414834)};
\draw [color=orange] plot [mark=*, mark size=1] coordinates{(-0.783793963663856, -0.108103018168072, 0.7745966692414834)};
\draw [color=orange] plot [mark=*, mark size=1] coordinates{(-0.108103018168072, -0.108103018168072, 0.7745966692414834)};
\path (-1,-1,-1) coordinate (A) (1,-1,-1) coordinate (B) (-1,1,-1) coordinate (C); 
\path (-1,-1,1) coordinate (D) (1,-1,1) coordinate (E) (-1,1,1) coordinate (F);
\draw (A)--(B)--(C)--(A)--(D)--(E)--(F)--(D);
\draw (B)--(E);
\draw (C)--(F);
\foreach \z in {-0.7745966692414834,0,0.7745966692414834}{
\draw [dashed, black, fill=black, fill opacity=0.075] (-1,-1,\z)--(1,-1,\z)--(-1,1,\z)--(-1,-1,\z);
}
\end{tikzpicture}
\qquad

\begin{tikzpicture}[x = {(0.75cm,0.25cm)}, y={(0.6cm,-0.15cm)}, z={(0cm,1cm)},]
\definecolor{myblue}{RGB}{79,193,232}
\definecolor{mygreen}{RGB}{160,213,104}
\draw plot [mark=*, mark size=1] coordinates{(-0.333333333333333, -0.333333333333333, -0.7745966692414834)};
\draw plot [mark=*, mark size=1] coordinates{(-0.888871894660414, 0.777743789320828, -0.7745966692414834)};
\draw plot [mark=*, mark size=1] coordinates{(0.777743789320828, -0.888871894660414, -0.7745966692414834)};
\draw plot [mark=*, mark size=1] coordinates{(-0.888871894660414, -0.888871894660414, -0.7745966692414834)};
\draw plot [mark=*, mark size=1] coordinates{(0.268421495491446, -0.408932576528214, -0.7745966692414834)};
\draw plot [mark=*, mark size=1] coordinates{(-0.859488918963232, -0.408932576528214, -0.7745966692414834)};
\draw plot [mark=*, mark size=1] coordinates{(-0.408932576528214, -0.859488918963232, -0.7745966692414834)};
\draw plot [mark=*, mark size=1] coordinates{(0.268421495491446, -0.859488918963232, -0.7745966692414834)};
\draw [color=blue!90] plot [mark=*, mark size=1] coordinates{(-0.859488918963232, 0.268421495491446, -0.7745966692414834)};
\draw plot [mark=*, mark size=1] coordinates{(-0.408932576528214, 0.268421495491446, -0.7745966692414834)};
\draw plot [mark=*, mark size=1] coordinates{(-0.333333333333333, -0.333333333333333, 0)};
\draw plot [mark=*, mark size=1] coordinates{(-0.888871894660414, 0.777743789320828, 0)};
\draw plot [mark=*, mark size=1] coordinates{(0.777743789320828, -0.888871894660414, 0)};
\draw plot [mark=*, mark size=1] coordinates{(-0.888871894660414, -0.888871894660414, 0)};
\draw plot [mark=*, mark size=1] coordinates{(0.268421495491446, -0.408932576528214, 0)};
\draw plot [mark=*, mark size=1] coordinates{(-0.859488918963232, -0.408932576528214, 0)};
\draw plot [mark=*, mark size=1] coordinates{(-0.408932576528214, -0.859488918963232, 0)};
\draw plot [mark=*, mark size=1] coordinates{(0.268421495491446, -0.859488918963232, 0)};
\draw [color=blue!90] plot [mark=*, mark size=1] coordinates{(-0.859488918963232, 0.268421495491446, 0)};
\draw plot [mark=*, mark size=1] coordinates{(-0.408932576528214, 0.268421495491446, 0)};
\draw plot [mark=*, mark size=1] coordinates{(-0.333333333333333, -0.333333333333333, 0.7745966692414834)};
\draw plot [mark=*, mark size=1] coordinates{(-0.888871894660414, 0.777743789320828, 0.7745966692414834)};
\draw plot [mark=*, mark size=1] coordinates{(0.777743789320828, -0.888871894660414, 0.7745966692414834)};
\draw plot [mark=*, mark size=1] coordinates{(-0.888871894660414, -0.888871894660414, 0.7745966692414834)};
\draw plot [mark=*, mark size=1] coordinates{(0.268421495491446, -0.408932576528214, 0.7745966692414834)};
\draw plot [mark=*, mark size=1] coordinates{(-0.859488918963232, -0.408932576528214, 0.7745966692414834)};
\draw plot [mark=*, mark size=1] coordinates{(-0.408932576528214, -0.859488918963232, 0.7745966692414834)};
\draw [color=orange] plot [mark=*, mark size=1] coordinates{(0.268421495491446, -0.859488918963232, 0.7745966692414834)};
\draw [color=blue!90] plot [mark=*, mark size=1] coordinates{(-0.859488918963232, 0.268421495491446, 0.7745966692414834)};
\draw plot [mark=*, mark size=1] coordinates{(-0.408932576528214, 0.268421495491446, 0.7745966692414834)};
\path (-1,-1,-1) coordinate (A) (1,-1,-1) coordinate (B) (-1,1,-1) coordinate (C); 
\path (-1,-1,1) coordinate (D) (1,-1,1) coordinate (E) (-1,1,1) coordinate (F);
\draw (A)--(B)--(C)--(A)--(D)--(E)--(F)--(D);
\draw (B)--(E);
\draw (C)--(F);
\foreach \z in {-0.7745966692414834,0,0.7745966692414834}{
\draw [dashed, black, fill=black, fill opacity=0.075] (-1,-1,\z)--(1,-1,\z)--(-1,1,\z)--(-1,-1,\z);
}
\end{tikzpicture}

\qquad

\begin{tikzpicture}[x = {(0.75cm,0.25cm)}, y={(0.6cm,-0.15cm)}, z={(0cm,1cm)},]
\definecolor{myblue}{RGB}{79,193,232}
\definecolor{mygreen}{RGB}{160,213,104}
\draw plot [mark=*, mark size=1] coordinates{(-0.333333333333333, -0.333333333333333, -0.8611363115940526)};
\draw plot [mark=*, mark size=1] coordinates{(-0.888871894660414, 0.777743789320828, -0.8611363115940526)};
\draw plot [mark=*, mark size=1] coordinates{(0.777743789320828, -0.888871894660414, -0.8611363115940526)};
\draw plot [mark=*, mark size=1] coordinates{(-0.888871894660414, -0.888871894660414, -0.8611363115940526)};
\draw plot [mark=*, mark size=1] coordinates{(0.268421495491446, -0.408932576528214, -0.8611363115940526)};
\draw plot [mark=*, mark size=1] coordinates{(-0.859488918963232, -0.408932576528214, -0.8611363115940526)};
\draw plot [mark=*, mark size=1] coordinates{(-0.408932576528214, -0.859488918963232, -0.8611363115940526)};
\draw plot [mark=*, mark size=1] coordinates{(0.268421495491446, -0.859488918963232, -0.8611363115940526)};
\draw plot [mark=*, mark size=1] coordinates{(-0.859488918963232, 0.268421495491446, -0.8611363115940526)};
\draw plot [mark=*, mark size=1] coordinates{(-0.408932576528214, 0.268421495491446, -0.8611363115940526)};
\draw plot [mark=*, mark size=1] coordinates{(-0.333333333333333, -0.333333333333333, -0.33998104358485626)};
\draw plot [mark=*, mark size=1] coordinates{(-0.888871894660414, 0.777743789320828, -0.33998104358485626)};
\draw plot [mark=*, mark size=1] coordinates{(0.777743789320828, -0.888871894660414, -0.33998104358485626)};
\draw plot [mark=*, mark size=1] coordinates{(-0.888871894660414, -0.888871894660414, -0.33998104358485626)};
\draw plot [mark=*, mark size=1] coordinates{(0.268421495491446, -0.408932576528214, -0.33998104358485626)};
\draw plot [mark=*, mark size=1] coordinates{(-0.859488918963232, -0.408932576528214, -0.33998104358485626)};
\draw plot [mark=*, mark size=1] coordinates{(-0.408932576528214, -0.859488918963232, -0.33998104358485626)};
\draw plot [mark=*, mark size=1] coordinates{(0.268421495491446, -0.859488918963232, -0.33998104358485626)};
\draw plot [mark=*, mark size=1] coordinates{(-0.859488918963232, 0.268421495491446, -0.33998104358485626)};
\draw plot [mark=*, mark size=1] coordinates{(-0.408932576528214, 0.268421495491446, -0.33998104358485626)};
\draw plot [mark=*, mark size=1] coordinates{(-0.333333333333333, -0.333333333333333, 0.33998104358485626)};
\draw plot [mark=*, mark size=1] coordinates{(-0.888871894660414, 0.777743789320828, 0.33998104358485626)};
\draw plot [mark=*, mark size=1] coordinates{(0.777743789320828, -0.888871894660414, 0.33998104358485626)};
\draw plot [mark=*, mark size=1] coordinates{(-0.888871894660414, -0.888871894660414, 0.33998104358485626)};
\draw plot [mark=*, mark size=1] coordinates{(0.268421495491446, -0.408932576528214, 0.33998104358485626)};
\draw plot [mark=*, mark size=1] coordinates{(-0.859488918963232, -0.408932576528214, 0.33998104358485626)};
\draw plot [mark=*, mark size=1] coordinates{(-0.408932576528214, -0.859488918963232, 0.33998104358485626)};
\draw plot [mark=*, mark size=1] coordinates{(0.268421495491446, -0.859488918963232, 0.33998104358485626)};
\draw [color=blue!90] plot [mark=*, mark size=1] coordinates{(-0.859488918963232, 0.268421495491446, 0.33998104358485626)};
\draw plot [mark=*, mark size=1] coordinates{(-0.408932576528214, 0.268421495491446, 0.33998104358485626)};
\draw plot [mark=*, mark size=1] coordinates{(-0.333333333333333, -0.333333333333333, 0.8611363115940526)};
\draw plot [mark=*, mark size=1] coordinates{(-0.888871894660414, 0.777743789320828, 0.8611363115940526)};
\draw plot [mark=*, mark size=1] coordinates{(0.777743789320828, -0.888871894660414, 0.8611363115940526)};
\draw plot [mark=*, mark size=1] coordinates{(-0.888871894660414, -0.888871894660414, 0.8611363115940526)};
\draw plot [mark=*, mark size=1] coordinates{(0.268421495491446, -0.408932576528214, 0.8611363115940526)};
\draw plot [mark=*, mark size=1] coordinates{(-0.859488918963232, -0.408932576528214, 0.8611363115940526)};
\draw plot [mark=*, mark size=1] coordinates{(-0.408932576528214, -0.859488918963232, 0.8611363115940526)};
\draw plot [mark=*, mark size=1] coordinates{(0.268421495491446, -0.859488918963232, 0.8611363115940526)};
\draw plot [mark=*, mark size=1] coordinates{(-0.859488918963232, 0.268421495491446, 0.8611363115940526)};
\draw plot [mark=*, mark size=1] coordinates{(-0.408932576528214, 0.268421495491446, 0.8611363115940526)};
\path (-1,-1,-1) coordinate (A) (1,-1,-1) coordinate (B) (-1,1,-1) coordinate (C); 
\path (-1,-1,1) coordinate (D) (1,-1,1) coordinate (E) (-1,1,1) coordinate (F);
\draw (A)--(B)--(C)--(A)--(D)--(E)--(F)--(D);
\draw (B)--(E);
\draw (C)--(F);
\foreach \z in {-0.8611363115940526,-0.33998104358485626,0.33998104358485626,0.8611363115940526}{
\draw [dashed, black, fill=black, fill opacity=0.075] (-1,-1,\z)--(1,-1,\z)--(-1,1,\z)--(-1,-1,\z);
}
\end{tikzpicture}
}
    \caption{An intermediate state shown in the middle is introduced to decompose interpolation from $p=2$ to $p=3$ into one block diagonal and one sparse factor.
First, an interpolation is carried out on triangular planes only resulting in a block diagonal operator as indicated with the orange points.
The next step is interpolating values along the z-direction that results in a sparse operator as indicated with the blue points.}
    \label{fig:priplanarint}
\end{figure}

However, a prism can be generated by a tensor production of a triangle and a line that is perpendicular to the triangle.
Depending on the polynomial order, there are a number of triangular planar regions inside a prism and the solution points are located on these regions.
Therefore, obtaining modal coefficients at quadrature points using the nodal coefficients at solution points, which is an operation carried out by the \textit{qptsu} kernel can be simplified and reformulated in a two part operation.
The first operation obtains the solution at the quadrature points within triangular planes in a prism, and then a second operation obtains the solution at a set of triangular planes inside the prism.
The procedure is illustrated in Figure \ref{fig:priplanarint}.
The former requires a block structured operator while the latter is a sparse operation.
As there is a sparse component in the factored formulation, cache blocking can be used effectively to implement this.

\section{Application of Cache Blocking in Tensor Product Factorisation}
\label{sec:cleverthing}
\begin{figure}[h!]
  \centering
  \includegraphics[width=\textwidth]{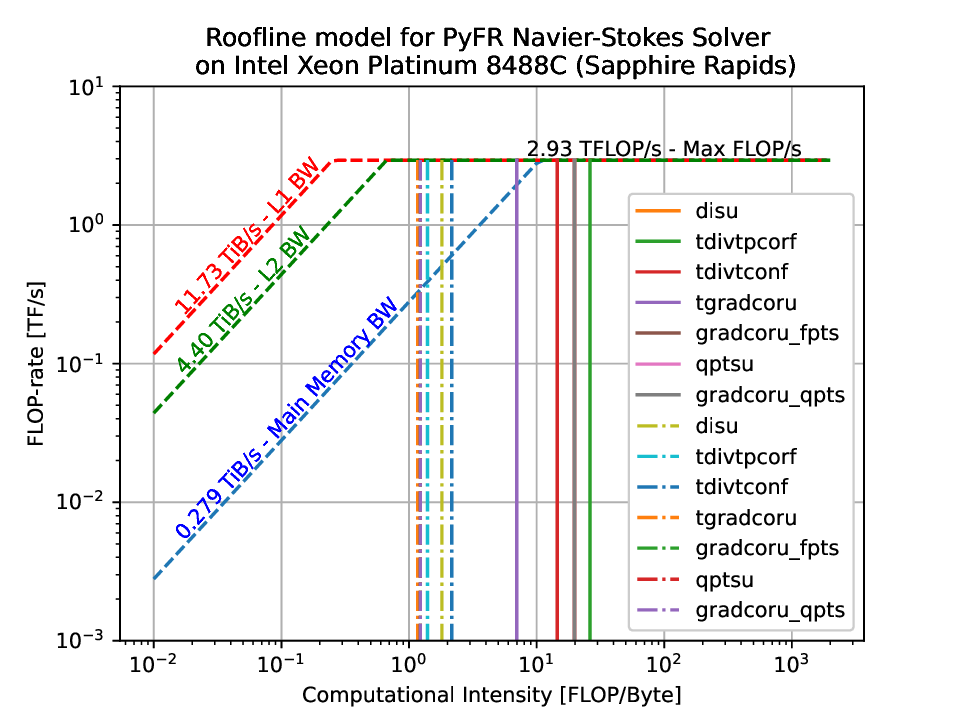}
  \caption{A selection of the Navier-Stokes kernels are shown in the figure for anti-aliasing off and full anti-aliasing on a full hexahedral mesh. The remaining kernels have similar properties if not exactly the same.}
  \label{fig:NSroofline}
\end{figure}

There are a few different ways to implement the tensor product operations after obtaining the sparse factorisation.
The naive implementation would simply call a matrix multiplication library for each sparse factor.
This would reduce the FLOP requirement significantly due to reduced number of non-zero entries in the sparse factors provided that a sparse matrix multiplication library is used.
However, it would result in increased bandwidth use because the intermediate solutions between each of the separate kernel invocations have to be stored in memory and this results in unnecessary reads and writes. 
The alternative is using cache blocking strategies.
These make it possible to store the intermediate results in CPU cache, and reduce the main memory access and data movement requirements significantly.
Using sparse factors rather than a single dense matrix reduces the FLOP requirement, and cache blocking strategy eliminates the excess data movement by utilizing CPU cache as an intermediate storage. 
Therefore, it is possible to get the best of both worlds by enabling cache blocking for the tensor product operations. 
Table \ref{table:smart} shows how the FLOP and data movement requirements differ between the regular approach where a single dense matrix is used, the naive approach where a matrix multiplication library is called for each sparse factor, and the ideal approach where cache blocking is enabled together with the sparse factors.
\begin{table}[h!]
\caption{FLOP and Data movement requirements for the \textit{tdivtconf} kernel on hexahedral elements for three different implementations}\label{table:smart}
\centering
\scalebox{1}{
\begin{tabular}{l||l|l|l}
 \multicolumn{1}{c}{} & \multicolumn{1}{c|}{ } & \multicolumn{2}{c}{Factored} \\ \cline{2-4}
          & Dense & Direct & Cache Blocked \\ \hline\hline
FLOP [GFLOPS]       &  30.00  &  4.50   & 4.50   \\ \hline
Data Movement [GiB] & \: 1.95  & 7.81  & 1.95  \\ \hline
FLOP/Byte           & 14.48 & 0.54  & 2.16
\end{tabular}
}
\end{table}
Furthermore, a roofline model \cite{roofline} demonstrating the FLOP/Byte ratio for a selection of Navier-Stokes kernels with full anti-aliasing are given in Figure \ref{fig:NSroofline} before and after applying tensor product factorisation using the cache blocking strategy.
It is clear from Table \ref{table:smart} and Figure \ref{fig:NSroofline} that the tensor product factorisation using the cache blocking strategy for the standalone tensor product kernels reduces the FLOP requirement while keeping the data movement requirement from main memory the same.
As a result, the profile of these individual kernels shifted towards a smaller FLOP/Byte region where they would normally be bandwidth bound from main memory.
However, now that most of the kernels are bandwidth bound, a kernel grouping strategy can be implemented to eliminate the unnecessary data movements between kernels that exchange data.
When these approaches are implemented together, there is a reduction in both FLOP and data movement requirements.

In summary, tensor product factorisation with cache blocking reduces the FLOP requirement significantly while keeping the amount of data movement from main memory the same.
Then, a kernel grouping strategy can be implemented to reduce the total amount of data movement requirements to improve the performance even further.

\section{Theoretical Predictions and Benchmarking in PyFR}\label{sec:grouping}
\begin{figure}[h!]
  \centering
  \includegraphics[width=\textwidth]{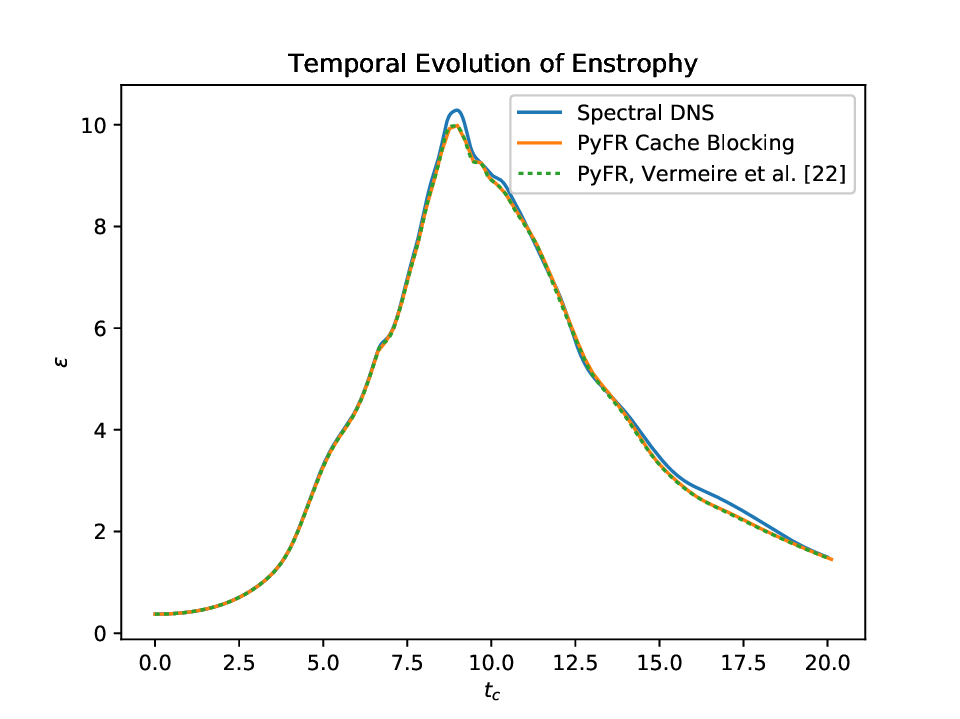}
  \caption{Temporal evolution of enstrophy obtained with cache blocking approach implemented in PyFR. Simulation is carried out on a $52^3$ all hexahedral mesh at $p=4$ with anti-aliasing off.}
  \label{fig:enstrophyNS}
\end{figure}

\begin{figure}[h!]
  \centering
  \includegraphics[width=\textwidth]{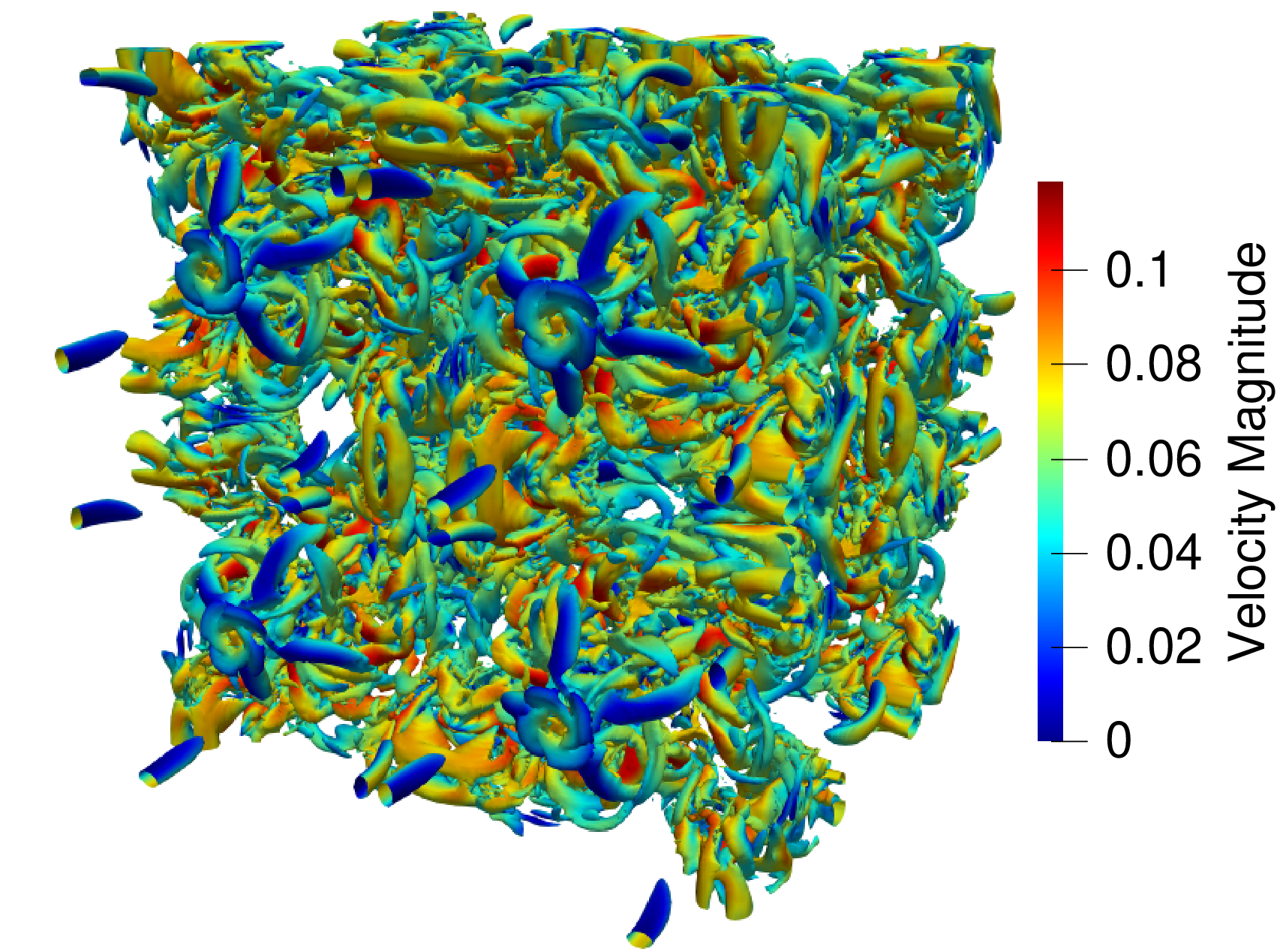}
  \caption{Q-criterion isosurfaces coloured based on velocity magnitude for a snapshot at $t=10t_c$.}
  \label{fig:qcritNS}
\end{figure}

A kernel grouping configuration for the Navier-Stokes solver is examined in this section both for anti-aliasing on.
The test case used for the performance comparisons is the compressible Taylor-Green Vortex.
It is a typical test case for the validation of high-order software and it is used in \cite{VERMEIRE2017497} for comparing PyFR performance against industry standard codes.
Also, it is one of the test cases in the high-order workshops \cite{HOReview}.
The configuration and mesh files used here are taken from Vermeire \cite{VERMEIRE2017497}, and the numerical results are compared against spectral DNS results from van Rees et al. \cite{VANREES20112794}.
Taylor-Green Vortex case is initialized with the following initial conditions
\begin{equation}
\begin{split}
v_x &= +U_0\sin(x/L)\cos(y/L)\cos(z/L),	\\
v_y &= +U_0\cos(x/L)\sin(y/L)\cos(z/L),	\\
v_z &= 0,	\\
p &= p_0 + \frac{\rho_0 U_0^2}{16}(\cos(2x/L) + \cos(2y/L))(\cos(2z/L)+2), \\
\rho &= \frac{p}{RT_0}
\end{split}
\end{equation}
in a cube domain with periodic boundary conditions in all the faces with dimensions $[-\pi L, +\pi L]^3$.
Performance comparisons of the cache blocking approach is executed at two different polynomial orders, $p=3$ and $p=4$, with a structured mesh consisting of $64^3$ and $52^3$ hexahedra respectively so that the degree of freedom is approximately equivalent.
The Reynolds number $Re = 1600$ based on the length $L$ and velocity $U_0$.
The Mach number based on $U_0$ is $M=0.1$ so that the simulation is effectively incompressible.
Firstly, a simulation is carried out using the cache blocking approach in order to validate the implementation. 
The evolution of enstropy is compared with the spectral DNS result \cite{VANREES20112794} and shown in Figure \ref{fig:enstrophyNS}.
The results from cache blocking show a good agreement with the spectral DNS results.
A snapshot at $t=10t_c$ is shown in Figure \ref{fig:qcritNS} with isosurfaces based on Q-criterion \cite{qcrit} and coloured by velocity magnitude.

\subsection{Kernel Grouping Configuration}\label{sec:NSgrouping}
\begin{figure}[p!]
    \centering
    \vspace{-1cm}
\begin{tikzpicture}

\def\yloc{0}
\def\name{negdivconf}
\node[draw, fill=red!50] (\name) at (0,\yloc) {\name};
\node (\name_l) at (-2,\yloc) {$\mathbf{R}^{(u)}$};
\node (\name_r) at (2,\yloc) {$\mathbf{R}^{(u)}$}; 
\node[draw,dotted,fit=(\name) (\name_l) (\name_r)] {};
\path[->, very thick] (\name_l) edge (\name);
\path[->, very thick] (\name) edge (\name_r);

\draw[->, very thick] (\name_r) to [out=0,in=-180] (4,\yloc);

\node (one) at (-5.75,0.75) {5};

\def\yloc{1.5}
\def\name{tdivtconf}
\node[draw, fill=blue!75!green!40] (\name) at (0,\yloc) {\name};
\node (\name_l) at (-2,\yloc) {$\mathbf{U}^{(f)}$};
\node (\name_r) at (2,\yloc) {$\mathbf{R}^{(u)}$}; 
\node[draw,dotted,fit=(\name) (\name_l) (\name_r)] {};
\path[->, very thick] (\name_l) edge (\name);
\path[->, very thick] (\name) edge (\name_r);

\draw[->, very thick, dashed, gray] (\name_r) to [out=0,in=-180] (4,\yloc);
\draw[->, very thick] (-5,\yloc) to [out=0,in=-180] (\name_l);

\draw[gray,dashed] (-4,\yloc+0.75) -- (4,\yloc+0.75);

\def\yloc{3}
\node[draw, fill=pink] (comm_flux) at (0,\yloc) {comm\_flux};
\node (comm_flux_ll) at (-3,\yloc) {$\mathbf{U}^{(f)}$,};
\node (comm_flux_l) at (-2,\yloc) {$\mathbf{F}^{(f)}$};
\node (comm_flux_r) at (2,\yloc) {$\mathbf{U}^{(f)}$}; 
\node[draw,dotted,fit=(comm_flux)(comm_flux_ll) (comm_flux_l) (comm_flux_r)] {};
\path[->, very thick] (comm_flux_l) edge (comm_flux);
\path[->, very thick] (comm_flux) edge (comm_flux_r);

\draw[->, very thick] (comm_flux_r) to [out=0,in=-180] (4,\yloc);
\draw[->, very thick] (-5,\yloc) to [out=0,in=-180] (comm_flux_ll);

\node (one) at (-5.75,\yloc) {4};
\draw[gray,dashed] (-4,\yloc+0.75) -- (4,\yloc+0.75);

\def\yloc{4.5}
\def\name{tdivpcorf}
\node[draw, fill=blue!75!green!40] (\name) at (0,\yloc) {\name};
\node (\name_l) at (-2,\yloc) {$\mathbf{F}^{(u)}$};
\node (\name_r) at (2,\yloc) {$\mathbf{R}^{(u)}$}; 
\node[draw,dotted,fit=(\name) (\name_l) (\name_r)] {};
\path[->, very thick] (\name_l) edge (\name);
\path[->, very thick] (\name) edge (\name_r);

\draw[->, very thick] (\name_r) to [out=0,in=-180] (4,\yloc);

\node (one) at (-5.75,9.75) {3};

\def\yloc{6}
\def\name{tdisf}
\node[draw, fill=red!50] (\name) at (0,\yloc) {\name};
\node (\name_ll) at (-3,\yloc) {$\mathbf{U}^{(q)}$,};
\node (\name_l) at (-2,\yloc) {$\mathbf{F}^{(q)}$};
\node (\name_r) at (2,\yloc) {$\mathbf{F}^{(q)}$}; 
\node[draw,dotted,fit=(\name) (\name_ll) (\name_l) (\name_r)] {};
\path[->, very thick] (\name_l) edge (\name);
\path[->, very thick] (\name) edge (\name_r);



\def\yloc{7.5}
\node[draw, fill=blue!75!green!40] (gradcoru_q) at (0,\yloc) {gradcoru\_qpts};
\node (gradcoru_q_l) at (-2,\yloc) {$\mathbf{F}^{(u)}$};
\node (gradcoru_q_r) at (2,\yloc) {$\mathbf{F}^{(q)}$}; 
\node[draw,dotted,fit=(gradcoru_q) (gradcoru_q_l) (gradcoru_q_r)] {};
\path[->, very thick] (gradcoru_q_l) edge (gradcoru_q);
\path[->, very thick] (gradcoru_q) edge (gradcoru_q_r);



\def\yloc{9}
\node[draw, fill=blue!75!green!40] (qptsu) at (0,\yloc) {qptsu};
\node (qptsu_l) at (-2,\yloc) {$\mathbf{U}^{(u)}$};
\node (qptsu_r) at (2,\yloc) {$\mathbf{U}^{(q)}$}; 
\node[draw,dotted,fit=(qptsu) (qptsu_l) (qptsu_r)] {};
\path[->, very thick] (qptsu_l) edge (qptsu);
\path[->, very thick] (qptsu) edge (qptsu_r);

\draw[->, very thick, looseness=0.5] (-3.75,15) to [out=0,in=-180] (qptsu_l);


\def\yloc{10.5}
\node[draw, fill=blue!75!green!40] (gradcoru_fpts) at (0,\yloc) {gradcoru\_fpts};
\node (gradcoru_fpts_l) at (-2,\yloc) {$\mathbf{F}^{(u)}$};
\node (gradcoru_fpts_r) at (2,\yloc) {$\mathbf{F}^{(f)}$}; 
\node[draw,dotted,fit=(gradcoru_fpts) (gradcoru_fpts_l) (gradcoru_fpts_r)] {};
\path[->, very thick] (gradcoru_fpts_l) edge (gradcoru_fpts);
\path[->, very thick] (gradcoru_fpts) edge (gradcoru_fpts_r);

\draw[->, very thick] (gradcoru_fpts_r) to [out=0,in=-180] (4,\yloc);


\def\yloc{12}
\def\name{gradcoru}
\node[draw, fill=red!50] (\name) at (0,\yloc) {\name};
\node (\name_l) at (-2,\yloc) {$\mathbf{F}^{(u)}$};
\node (\name_r) at (2,\yloc) {$\mathbf{F}^{(u)}$}; 
\node[draw,dotted,fit=(\name) (\name_l) (\name_r)] {};
\path[->, very thick] (\name_l) edge (\name);
\path[->, very thick] (\name) edge (\name_r);



\def\yloc{13.5}
\def\name{tgradcoru}
\node[draw, fill=blue!75!green!40] (\name) at (0,\yloc) {\name};
\node (\name_l) at (-2,\yloc) {$\mathbf{U}^{(f)}$};
\node (\name_r) at (2,\yloc) {$\mathbf{F}^{(u)}$}; 
\node[draw,dotted,fit=(\name) (\name_l) (\name_r)] {};
\path[->, very thick] (\name_l) edge (\name);
\path[->, very thick] (\name) edge (\name_r);

\draw[->, very thick] (-5,\yloc) to [out=0,in=-180] (\name_l);


\def\yloc{15}
\def\name{tgradpcoru}
\node[draw, fill=blue!75!green!40] (\name) at (0,\yloc) {\name};
\node (\name_l) at (-2,\yloc) {$\mathbf{U}^{(u)}$};
\node (\name_r) at (2,\yloc) {$\mathbf{F}^{(u)}$}; 
\node[draw,dotted,fit=(\name) (\name_l) (\name_r)] {};
\path[->, very thick] (\name_l) edge (\name);
\path[->, very thick] (\name) edge (\name_r);

\draw[->, very thick] (-5,\yloc) to [out=0,in=-180] (\name_l);

\draw[gray,dashed] (-4,\yloc+0.75) -- (4,\yloc+0.75);

\def\yloc{16.5}
\def\name{conu}
\node[draw, fill=pink] (con_u) at (0,\yloc) {con\_u};
\node (con_u_l) at (-2,\yloc) {$\mathbf{U}^{(f)}$};
\node (con_u_r) at (2,\yloc) {$\mathbf{U}^{(f)}$}; 
\node[draw,dotted,fit=(con_u) (con_u_l) (con_u_r)] {};
\path[->, very thick] (con_u_l) edge (con_u);
\path[->, very thick] (con_u) edge (con_u_r);

\draw[->, very thick] (con_u_r) to [out=0,in=-180] (4,\yloc);
\draw[->, very thick] (-5,\yloc) to [out=0,in=-180] (con_u_l);

\node (one) at (-5.75,\yloc) {2};
\draw[gray,dashed] (-4,\yloc+0.75) -- (4,\yloc+0.75);

\def\yloc{18}
\def\name{disu}
\node[draw, fill=blue!75!green!40] (\name) at (0,\yloc) {\name};
\node (\name_l) at (-2,\yloc) {$\mathbf{U}^{(u)}$};
\node (\name_r) at (2,\yloc) {$\mathbf{U}^{(f)}$}; 
\node[draw,dotted,fit=(\name) (\name_l) (\name_r)] {};
\path[->, very thick] (\name_l) edge (\name);
\path[->, very thick] (\name) edge (\name_r);

\draw[->, very thick] (\name_r) to [out=0,in=-180] (4,\yloc);
\draw[->, very thick] (-5,\yloc) to [out=0,in=-180] (\name_l);

\node (one) at (-5.75,\yloc) {1};






\draw[->, very thick, looseness=0.75] (tdivtconf_r) to [out=-90,in=90] (negdivconf_l);

\draw[->, very thick, looseness=0.75] (tdisf_r) to [out=-90,in=90] (tdivpcorf_l);

\draw[->, very thick, looseness=0.75] (tgradcoru_r) to [out=-90,in=90] (gradcoru_l);
\draw[->, very thick, looseness=0.5] (tgradpcoru_r) .. controls(1,14) and (-5,15.5) .. (gradcoru_l); 


\draw[->, very thick, looseness=0.75] (gradcoru_r) to [out=-90,in=90] (gradcoru_fpts_l);

\draw[->, very thick, looseness=1] (gradcoru_r.south west) .. controls(1,11) and (-6,13) .. (gradcoru_q_l.west);


\draw[->, very thick, looseness=0.5] (qptsu_r) .. controls(1,8) and (-5,9.5) .. (tdisf_ll); 

\draw[->, very thick, looseness=0.75] (gradcoru_q_r) to [out=-90,in=90] (tdisf_l);

\definecolor{mygreen}{RGB}{160,213,104}
\draw[mygreen, very thick] (-4.25,-0.5) -- (-4.25,18.5);
\draw[mygreen, very thick] (3.25,-0.5) -- (3.25,18.5);

\node[draw=red,inner sep=8pt,
      fit=(tdivtconf) (tdivtconf_l) (tdivtconf_r)
          (negdivconf) (negdivconf_l) (negdivconf_r)] {};

\node[draw=red,inner sep=8pt,
      fit=(tgradpcoru) (tgradpcoru_l) (tgradpcoru_r)
          (tgradcoru) (tgradcoru_l) (tgradcoru_r)
          (gradcoru) (gradcoru_l) (gradcoru_r)
          (gradcoru_fpts) (gradcoru_fpts_l) (gradcoru_fpts_r)
          (tdisf) (tdisf_l) (tdisf_ll) (tdisf_r)
          (tdivpcorf) (tdivpcorf_l) (tdivpcorf_r)] {};

\end{tikzpicture}
    \caption{Kernel grouping configuration of the Navier-Stokes solver with flux anti-aliasing. Arrows going over green line indicate main memory access. Dashed gray arrow indicates write orders avoided due to cache blocking, while solid arrow indicate a definite read/write operation. Constant arrays such as $x$, which are not changed by the kernels, are omitted for simplicity.}
    \label{fig:NSAAKG}
\end{figure}
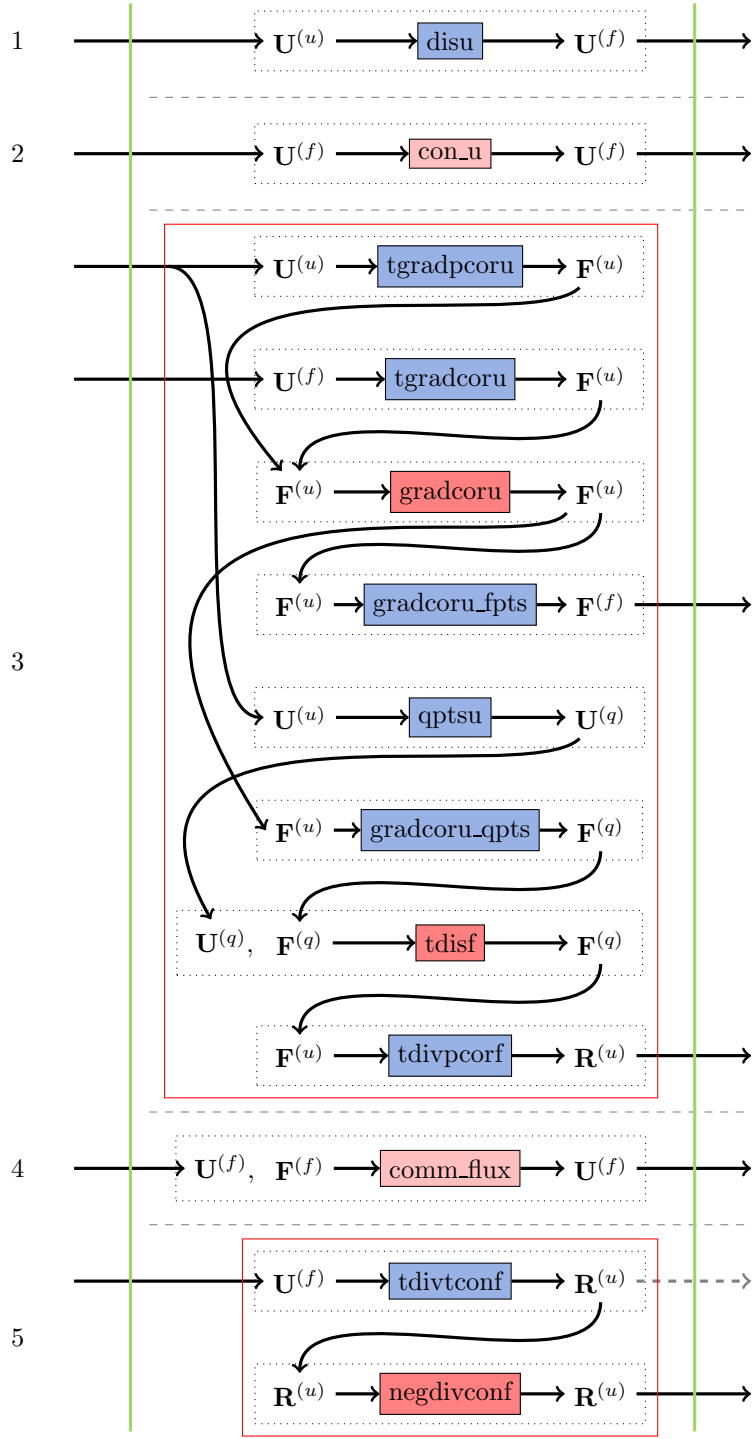

The main consideration when forming kernel groupings is the amount of bandwidth saving.
Bandwidth saving is maximised when a chain of kernels that write and read from the same location in the memory grouped together.
The more data kernels in a group exchange by writing and reading the same piece of memory the better the savings and therefore speedups.
Therefore the first step is analysing the kernels in the Navier-Stokes solver and determine chain of kernels that exchange data.
When compared with the Euler solver, there are more constraints in Navier-Stokes solver and this limits the number of possible kernel grouping configurations.
As with the Euler solver, eliminating the intermediate storage array $\mathbf{F}^{(u)}$ is possible in Navier-Stokes solver as well and it saves a considerable amount of bandwidth.
There are 6 kernels in total that operate on $\mathbf{F}^{(u)}$, and a kernel group with all of them can save a significant amount of data movement. 
The kernels are \textit{tgradpcoru}, \textit{tgradcoru}, \textit{gradcoru}, \textit{gradcoru\_fpts}, \textit{tdisf}, and \textit{tdivpcorf}.
In case of flux anti-aliasing the additional kernels, \textit{qptsu} and \textit{gradcoru\_qpts}, are also added to this kernel grouping as they also make use of $\mathbf{F}^{(u)}$ as well as introducing a new temporary storages $\mathbf{U}^{(q)}$ and $\mathbf{F}^{(q)}$ that can be stored entirely in the CPU cache for a single block.
Afterwards, there is only one other kernel grouping possibility is left which consist of \textit{tdivtconf} and \textit{negdivconf}.
A schematic of the kernel grouping configuration is given in Figure \ref{fig:NSAAKG} for full anti-aliasing.
Data movement requirements of the present kernel grouping configuration with anti-aliasing on are tabulated in Table \ref{table:NSKGBW}.
Further, the speedups obtained in three different platforms with an Arm based AWS Graviton3, Intel Xeon Platinum 8488C, and AMD EPYC 7742 are given in Table \ref{table:NSSpeedup}.
The highest speedup is 3.67x compared to PyFR v1.11.0 and it is obtained when running at $p=4$ on hexahedral elements with full anti-aliasing on Intel Xeon Platinum 8488C CPUs.
GDOF/s parameters are also tabulated in Table \ref{table:NSGDOFs} and compared with PyFR v1.11.0 and ZEFR \cite{ZEFR} on NVIDIA V100 GPU.
'GDOF/s' parameter makes it possible to compare performance of PyFR on different platforms such as GPUs and CPUs.
When cache blocking is enabled with anti-aliasing, single socket CPU performance of PyFR exceeds the single device GPU performance.

\begin{table}[h!]
\caption{Theoretical speedup predictions compared with observed speedups on AMD EPYC 7742, Arm based AWS Graviton3, and Intel Xeon Platinum 8488C CPUs.}\label{table:NSSpeedup}
\centering
\scalebox{0.98}{
\begin{tabular}{r||r|r|r|r|r|r|r|r}
          \multicolumn{1}{c}{} & \multicolumn{4}{c|}{AA off} & \multicolumn{4}{|c}{Full AA} \\ \cline{2-9}
          \multicolumn{1}{c}{} & \multicolumn{2}{c|}{Hexa} & \multicolumn{2}{|c|}{Prism} & \multicolumn{2}{|c|}{Hexa} & \multicolumn{2}{|c}{Prism} \\ \cline{2-9}

Speedups & $p=3$ & $p=4$ & $p=3$ & $p=4$ & $p=3$ & $p=4$ & $p=3$ & $p=4$        \\ \hline \hline
\textbf{Prediction} & \textbf{2.10} & \textbf{2.31} & \textbf{1.99} & \textbf{2.16} & \textbf{2.34} & \textbf{2.62} & \textbf{2.19} & \textbf{2.41} \\ 
AMD &        2.22 & 2.17 & 1.87 & 1.91 & 3.31 & 3.42 & 1.83 & 2.02 \\ 
Arm &        2.17 & 2.53 & 1.76 & 1.78 & 2.82 & 2.03 & 1.68 & 1.67 \\
Intel &      2.61 & 2.76 & 1.83 & 1.90 & 2.83 & 3.67 & 1.95 & 2.08 \\
\end{tabular}
}
\end{table}

\begin{table}[h!]
\caption{Comparison of GDOF/s: PyFR with Cache Blocking on AMD EPYC 7742 CPU, Arm based AWS Graviton3 CPU, Intel Xeon Platinum 8488C CPU, PyFR v1.11.0 on NVIDIA V100 GPU, and ZEFR on NVIDIA V100 GPU.}\label{table:NSGDOFs}
\centering
\scalebox{0.8}{
\begin{tabular}{r||r|r|r|r|r|r|r|r}

          \multicolumn{1}{c}{} & \multicolumn{4}{c}{AA off} & \multicolumn{4}{|c}{Full AA} \\ \cline{2-9}
          \multicolumn{1}{c}{} & \multicolumn{2}{c|}{Hexa} & \multicolumn{2}{|c|}{Prism} & \multicolumn{2}{|c|}{Hexa} & \multicolumn{2}{|c}{Prism} \\ \cline{2-9}

GDOF/s & $p=3$ & $p=4$ & $p=3$ & $p=4$ & $p=3$ & $p=4$ & $p=3$ & $p=4$        \\ \hline \hline
PyFR Cache Blocking - AMD   & 0.520 & 0.629 & 0.427 & 0.486 & 0.318 & 0.368 & 0.249 & 0.277\\ 
PyFR Cache Blocking - Arm   & 0.921 & 1.162 & 0.617 & 0.702 & 0.538 & 0.339 & 0.347 & 0.359 \\
PyFR Cache Blocking - Intel & 0.830 & 0.983 & 0.550 & 0.632 & 0.499 & 0.596 & 0.310 & 0.348 \\
PyFR v1.11.0 - NVIDIA V100  & 0.621 & 0.665 & & & 0.147 & 0.117 & &        \\  
ZEFR \cite{ZEFR} - NVIDIA V100 & 0.60 & 0.65 & - & - & - & -  & - & -      \\ 
\end{tabular}
}
\end{table}

\section{Conclusion}\label{sec:conclusion}
In this article, a methodology for applying cache blocking in FR has been extended and implemented for the Navier-Stokes solver with full anti-aliasing support on mixed grids in PyFR for CPUs.
Cache blocking was used to eliminate unnecessary data movements between kernels that exchange data at the main memory level, as an alternative to kernel fusion.
In particular, per-core private data cache on CPUs was used as a temporary storage to enable data exchanges between consecutive kernels.
As the amount of cache is limited, these kernels are executed on small sub-regions of the domain that fit in the available cache.
In addition to eliminating data exchanges between kernels, cache blocking was used to implement dense interpolation kernels associated with anti-aliasing as well.
These dense interpolation kernels first decomposed into sparse matrix factors using tensor product factorisation, and cache blocking is used as a temporary storage for the intermediate results when applying these chain of sparse matrix factors consecutively.
As a result, cache blocking saved a significant amount of data movement from main memory.
The amount of savings in data movements then used to construct a theoretical model to predict the performance gains.
Finally, the implementation was benchmarked using a compressible 3D Taylor-Green vortex test case on both hexahedral and prismatic grids, with third- and forth-order solution polynomials. 
A comparison between the theoretical expectations and speedups obtained in practice compared to PyFR v1.11.0 were given.

\section*{Acknowledgements}
The first author gratefully acknowledges the President’s PhD Scholarship provided by Imperial College London. The authors would like to thank the Engineering and Physical Sciences Research Council for their support via an Early Career Fellowship (EP/R030340/1).

\appendix
\section{Data Movement Requirements}
A breakdown of data movement requirements for PyFR v1.11.0 and PyFR with kernel grouping are provided in Tables \ref{table:NSBW} and \ref{table:NSKGBW}.
\begin{landscape}

\begin{table}[h!]
\caption{Current Navier-Stokes Data Movement Requirements.}\label{table:NSBW}
\centering
\scalebox{0.9}{
\begin{tabular}{l||c|c||r|r|r|r|r|r|r|r|r}
         \multicolumn{3}{r}{} & \multicolumn{9}{c}{Data Movement Requirements Per RHS Per Element [KiB/RHS/element]} \\ \cline{4-12}
         \multicolumn{3}{r}{} & \multicolumn{1}{c}{} & \multicolumn{4}{c|}{AA off} & \multicolumn{4}{|c}{Full AA} \\ \cline{5-12}
         \multicolumn{3}{r}{} & \multicolumn{1}{c}{} & \multicolumn{2}{c|}{Hexa} & \multicolumn{2}{|c|}{Prism} & \multicolumn{2}{|c|}{Hexa} & \multicolumn{2}{|c}{Prism} \\ \cline{5-12}
Kernel   & Input               & Output              & Formula & $p=3$ & $p=4$ & $p=3$ & $p=4$ & $p=3$ & $p=4$ & $p=3$ & $p=4$       \\ \hline \hline
\textit{disu}                        & $\mathbf{U}^\mathit{(u)}_e$ & $\mathbf{U}^\mathit{(f)}_e$ & $\mathcal{S}(\mathbf{U}^\mathit{(u)}_e)+\mathcal{S}(\mathbf{U}^\mathit{(f)}_e)$ & 6.25 & 10.74 & 4.22 & 7.03 & 8.36 & 13.32 & 5.66 & 8.79 \\ 
\textit{con\_u}                  & $\mathbf{D}^\mathit{(f)}_e$,$\mathbf{n}^\mathit{(f)}_e$ & $\mathbf{D}^\mathit{(f)}_e$ & $\mathcal{S}(\mathbf{n}^\mathit{(u)}_e)+2\mathcal{S}(\mathbf{D}^\mathit{(f)}_e)$ & 10.50 & 16.41 & 7.44 & 11.48 & 14.72 & 21.56 & 10.33 & 15.00 \\ 
\textit{tgradpcoru}                  & $\mathbf{U}^\mathit{(u)}_e$ & $\mathbf{F}^\mathit{(u)}_e$ & $\mathcal{S}(\mathbf{U}^\mathit{(u)}_e)+\mathcal{S}(\mathbf{F}^\mathit{(u)}_e)$ & 10.00 & 19.53 & 6.25 & 11.72 & 10.00 & 19.53 & 6.25 & 11.72 \\ 
\textit{tgradcoru}                  & $\mathbf{U}^\mathit{(f)}_e$ & $\mathbf{F}^\mathit{(u)}_e$ & $\mathcal{S}(\mathbf{U}^\mathit{(f)}_e)+2\mathcal{S}(\mathbf{F}^\mathit{(u)}_e)$ & 18.75 & 35.16 & 12.03 & 21.68 & 20.86 & 37.73 & 13.48 & 23.44 \\ 
\textit{gradcoru}                  & $\mathbf{F}^\mathit{(u)}_e$ , $\mathbf{x}^\mathit{(u)}_e$, $\mathbf{J}^\mathit{(u)}_e$& $\mathbf{F}^\mathit{(u)}_e$ & $\mathcal{S}(\mathbf{J}^\mathit{(u)}_e)+\mathcal{S}(\mathbf{x}^\mathit{(u)}_e)+2\mathcal{S}(\mathbf{F}^\mathit{(u)}_e)$ & 15.69 & 30.46 & 9.88 & 18.30 & 15.69 & 30.46 & 9.83 & 18.30 \\ 
\textit{gradcoru\_fpts}                  & $\mathbf{F}^\mathit{(u)}_e$ & $\mathbf{F}^\mathit{(f)}_e$ & $\mathcal{S}(\mathbf{F}^\mathit{(u)}_e)+\mathcal{S}(\mathbf{F}^\mathit{(f)}_e)$ & 18.75 & 49.80 & 20.63 & 21.09 & 25.08 & 39.96 & 16.99 & 26.37 \\ 
\textit{uqpts} & $\mathbf{U}^\mathit{(u)}_e$ & $\mathbf{U}^\mathit{(q)}_e$ & $\mathcal{S}(\mathbf{U}^\mathit{(u)}_e)+\mathcal{S}(\mathbf{U}^\mathit{(q)}_e)$ &
- & - & - & - & 7.38 & 13.32 & 4.49 & 7.85 \\ 
\textit{gradcoru\_qpts} & $\mathbf{F}^\mathit{(u)}_e$ & $\mathbf{F}^\mathit{(q)}_e$ & $\mathcal{S}(\mathbf{F}^\mathit{(u)}_e)+\mathcal{S}(\mathbf{F}^\mathit{(q)}_e)$ &
- & - & - & - & 22.15 & 39.96 & 13.48 & 23.55 \\ 
\textit{tdisf}                       & $\mathbf{U}^\mathit{(u)}_e$, $\mathbf{x}^\mathit{(u)}_e$, $\mathbf{F}^\mathit{(u)}_e$ & $\mathbf{F}^\mathit{(u)}_e$ & $\mathcal{S}(\mathbf{U}^\mathit{(u)}_e)+\mathcal{S}(\mathbf{x}^\mathit{(u)}_e)+2\mathcal{S}(\mathbf{F}^\mathit{(u)}_e)$ & 17.69 & 34.37 & 11.13 & 20.65 & 34.37 & 59.25 & 20.65 & 34.59 \\ 
\textit{tdivpcorf}                  & $\mathbf{F}^\mathit{(u)}_e$ & $\mathbf{R}^\mathit{(u)}_e$ & $\mathcal{S}(\mathbf{F}^\mathit{(u)}_e)+\mathcal{S}(\mathbf{R}^\mathit{(u)}_e)$ & 10.00 & 19.53 & 6.25 & 11.72 & 17.15 & 30.20 & 10.35 & 17.70 \\ 
\textit{comm\_flux}                  & $\mathbf{D}^\mathit{(f)}_e$,$\mathbf{F}^\mathit{(f)}_e$,$\mathbf{n}^\mathit{(f)}_e$ & $\mathbf{D}^\mathit{(f)}_e$ & $\mathcal{S}(\mathbf{n}^\mathit{(u)}_e)+\mathcal{S}(\mathbf{F}^\mathit{(f)}_e)+2\mathcal{S}(\mathbf{D}^\mathit{(f)}_e)$ & 21.75 & 33.98 & 15.40 & 23.79 & 32.30 & 46.88 & 22.63 & 32.58 \\ 
\textit{tdivtconf}                   & $\mathbf{U}^\mathit{(f)}_e$       & $\mathbf{R}^\mathit{(u)}_e$ & $\mathcal{S}(\mathbf{U}^\mathit{(f)}_e)+2\mathcal{S}(\mathbf{R}^\mathit{(u)}_e)$ & 8.75 & 15.63 & 5.78 & 9.96 & 10.85 & 18.20 & 7.23 & 11.72 \\ 
\textit{negdivconf}                  & $\mathbf{R}^\mathit{(u)}_e$,$\mathbf{J}^\mathit{(u)}_e$ & $\mathbf{R}^\mathit{(u)}_e$ & $\mathcal{S}(\mathbf{J}^\mathit{(u)}_e)+2\mathcal{S}(\mathbf{R}^\mathit{(u)}_e)$ & 5.50 & 10.74 & 3.44 & 6.45 & 5.50 & 10.74 & 3.44 & 6.45 \\ \cline{2-12}
         \multicolumn{4}{r|}{Total Bandwidth}      & 143.63 & 276.35 & 102.44 & 163.88 & 224.41 & 381.12 & 144.80 & 238.05 \\  
\end{tabular}
}

\end{table}
\end{landscape}

\begin{landscape}

\begin{table}[p!]
\caption{Navier-Stokes Data Movement Requirements with kernel grouping.}\label{table:NSKGBW}
\centering
\scalebox{0.9}{
\begin{tabular}{l||c|c||r|r|r|r|r|r|r|r|r}
         \multicolumn{3}{r}{} & \multicolumn{9}{c}{Data Movement Requirements Per RHS Per Element [KiB/RHS/element]} \\ \cline{4-12}
         \multicolumn{3}{r}{} & \multicolumn{1}{c}{} & \multicolumn{4}{c|}{AA off} & \multicolumn{4}{|c}{Full AA} \\ \cline{5-12}
         \multicolumn{3}{r}{} & \multicolumn{1}{c}{} & \multicolumn{2}{c|}{Hexa} & \multicolumn{2}{|c|}{Prism} & \multicolumn{2}{|c|}{Hexa} & \multicolumn{2}{|c}{Prism} \\ \cline{5-12}
Kernel   & Input               & Output              & Formula & $p=3$ & $p=4$ & $p=3$ & $p=4$ & $p=3$ & $p=4$ & $p=3$ & $p=4$       \\ \hline \hline
\textit{disu}                        & $\mathbf{U}^\mathit{(u)}_e$ & $\mathbf{U}^\mathit{(f)}_e$ & $\mathcal{S}(\mathbf{U}^\mathit{(u)}_e)+\mathcal{S}(\mathbf{U}^\mathit{(f)}_e)$ & 6.25 & 10.74 & 4.22 & 7.03 & 8.36 & 13.32 & 5.66 & 8.79 \\ 
\textit{con\_u}                  & $\mathbf{D}^\mathit{(f)}_e$,$\mathbf{n}^\mathit{(f)}_e$ & $\mathbf{D}^\mathit{(f)}_e$ & $\mathcal{S}(\mathbf{n}^\mathit{(u)}_e)+2\mathcal{S}(\mathbf{D}^\mathit{(f)}_e)$ & 10.50 & 16.41 & 7.44 & 11.48 & 14.72 & 21.56 & 10.33 & 15.00 \\ 
\textit{tgradpcoru}                  & $\mathbf{U}^\mathit{(u)}_e$ & $\mathbf{F}^\mathit{(u)}_e$ & $\mathcal{S}(\mathbf{U}^\mathit{(u)}_e)$ & 2.50 & 4.88 & 1.56 & 2.93 & 2.50 & 4.88 & 1.56 & 2.93 \\ 
\textit{tgradcoru}                  & $\mathbf{U}^\mathit{(f)}_e$ & $\mathbf{F}^\mathit{(u)}_e$ & $\mathcal{S}(\mathbf{U}^\mathit{(f)}_e)$ & 3.75 & 5.86 & 2.66 & 4.10 & 5.86 & 8.44 & 4.10 & 5.86 \\ 
\textit{gradcoru}                  & $\mathbf{F}^\mathit{(u)}_e$ , $\mathbf{x}^\mathit{(c)}_e$, $\mathbf{J}^\mathit{(u)}_e$& $\mathbf{F}^\mathit{(u)}_e$ & $\mathcal{S}(\mathbf{J}^\mathit{(u)}_e)+\mathcal{S}(\mathbf{x}^\mathit{(u)}_e)$ & 0.69 & 1.16 & 0.45 & 0.73 & 0.69 & 1.16 & 0.45 & 0.73 \\ 
\textit{gradcoru\_fpts}                  & $\mathbf{F}^\mathit{(u)}_e$ & $\mathbf{F}^\mathit{(f)}_e$ & $\mathcal{S}(\mathbf{F}^\mathit{(f)}_e)$ & 11.25 & 17.58 & 7.97 & 12.30 & 17.58 & 25.31 & 12.30 & 17.58 \\ 
\textit{uqpts} & $\mathbf{U}^\mathit{(u)}_e$ & $\mathbf{U}^\mathit{(q)}_e$ & 0 & - & - & - & - & 0 & 0 & 0 & 0 \\ 
\textit{gradcoru\_qpts} & $\mathbf{F}^\mathit{(u)}_e$ & $\mathbf{F}^\mathit{(q)}_e$ & 0 & - & - & - & - & 0 & 0 & 0 & 0 \\ 
\textit{tdisf}                       & $\mathbf{U}^\mathit{(u)}_e$, $\mathbf{x}^\mathit{(c)}_e$, $\mathbf{F}^\mathit{(u)}_e$ & $\mathbf{F}^\mathit{(u)}_e$ & $\mathcal{S}(\mathbf{x}^\mathit{(u)}_e)$ & 0.19 & 0.19 & 0.14 & 0.14 & 0.19 & 0.19 & 0.14 & 0.14 \\ 
\textit{tdivpcorf}                  & $\mathbf{F}^\mathit{(u)}_e$ & $\mathbf{R}^\mathit{(u)}_e$ & $\mathcal{S}(\mathbf{R}^\mathit{(u)}_e)$ & 2.50 & 4.88 & 1.56 & 2.93 & 2.50 & 4.88 & 1.56 & 2.93 \\ 
\textit{comm\_flux}                  & $\mathbf{D}^\mathit{(f)}_e$,$\mathbf{F}^\mathit{(f)}_e$,$\mathbf{n}^\mathit{(f)}_e$ & $\mathbf{D}^\mathit{(f)}_e$ & $\mathcal{S}(\mathbf{n}^\mathit{(u)}_e)+\mathcal{S}(\mathbf{F}^\mathit{(f)}_e)+2\mathcal{S}(\mathbf{D}^\mathit{(f)}_e)$ & 21.75 & 33.98 & 15.41 & 23.79 & 32.30 & 46.88 & 22.63 & 32.58 \\ 
\textit{tdivtconf}                   & $\mathbf{U}^\mathit{(f)}_e$       & $\mathbf{R}^\mathit{(u)}_e$ & $\mathcal{S}(\mathbf{U}^\mathit{(f)}_e)+\mathcal{S}(\mathbf{R}^\mathit{(u)}_e)$ & 6.25 & 10.74 & 4.22 & 7.03 & 8.36 & 13.32 & 5.66 & 8.79 \\ 
\textit{negdivconf}                  & $\mathbf{R}^\mathit{(u)}_e$,$\mathbf{J}^\mathit{(u)}_e$ & $\mathbf{R}^\mathit{(u)}_e$ & $\mathcal{S}(\mathbf{J}^\mathit{(u)}_e)+\mathcal{S}(\mathbf{R}^\mathit{(u)}_e)$ & 3.00 & 5.86 & 1.88 & 3.52 & 3.00 & 5.86 & 1.88 & 3.52 \\ \cline{2-12}
         \multicolumn{4}{r||}{Total}      & 68.43 & 112.10 & 47.36 & 75.84 & 95.86 & 145.62 & 66.15 & 98.70 \\ \cline{4-12}         
         \multicolumn{4}{r||}{Predicted Speedup}      & 2.10 & 2.31 & 1.99 & 2.16 & 2.34 & 2.62 & 2.19 & 2.41 \\
\end{tabular}
}

\end{table}
\end{landscape}

\section*{References}
\bibliography{mybibfile}

\begin{thebibliography}{10}
\expandafter\ifx\csname url\endcsname\relax
  \def\url#1{\texttt{#1}}\fi
\expandafter\ifx\csname urlprefix\endcsname\relax\def\urlprefix{URL }\fi
\expandafter\ifx\csname href\endcsname\relax
  \def\href#1#2{#2} \def\path#1{#1}\fi

\bibitem{Akkurt}
S.~Akkurt, F.~Witherden, P.~Vincent, Cache blocking strategies applied to flux
  reconstruction, Computer Physics Communications 271 (2022) 108193.
\newblock \href {http://dx.doi.org/https://doi.org/10.1016/j.cpc.2021.108193}
  {\path{doi:https://doi.org/10.1016/j.cpc.2021.108193}}.

\bibitem{Park17}
J.~S. Park, F.~D. Witherden, P.~E. Vincent,
  \href{https://doi.org/10.2514/1.J055304}{High-order implicit large-eddy
  simulations of flow over a naca0021 aerofoil}, AIAA Journal 55~(7) (2017)
  2186--2197.
\newblock \href {http://arxiv.org/abs/https://doi.org/10.2514/1.J055304}
  {\path{arXiv:https://doi.org/10.2514/1.J055304}}, \href
  {http://dx.doi.org/10.2514/1.J055304} {\path{doi:10.2514/1.J055304}}.
\newline\urlprefix\url{https://doi.org/10.2514/1.J055304}

\bibitem{ORSZAG198070}
S.~A. Orszag,
  \href{https://www.sciencedirect.com/science/article/pii/0021999180900054}{Spectral
  methods for problems in complex geometries}, Journal of Computational Physics
  37~(1) (1980) 70--92.
\newblock \href
  {http://dx.doi.org/https://doi.org/10.1016/0021-9991(80)90005-4}
  {\path{doi:https://doi.org/10.1016/0021-9991(80)90005-4}}.
\newline\urlprefix\url{https://www.sciencedirect.com/science/article/pii/0021999180900054}

\bibitem{Sherwin95}
S.~J. Sherwin, G.~E. Karniadakis,
  \href{https://onlinelibrary.wiley.com/doi/abs/10.1002/nme.1620382204}{A new
  triangular and tetrahedral basis for high-order (hp) finite element methods},
  International Journal for Numerical Methods in Engineering 38~(22) (1995)
  3775--3802.
\newblock \href
  {http://arxiv.org/abs/https://onlinelibrary.wiley.com/doi/pdf/10.1002/nme.1620382204}
  {\path{arXiv:https://onlinelibrary.wiley.com/doi/pdf/10.1002/nme.1620382204}},
  \href {http://dx.doi.org/https://doi.org/10.1002/nme.1620382204}
  {\path{doi:https://doi.org/10.1002/nme.1620382204}}.
\newline\urlprefix\url{https://onlinelibrary.wiley.com/doi/abs/10.1002/nme.1620382204}

\bibitem{Cantwell11}
C.~D. Cantwell, S.~J. Sherwin, R.~M. Kirby, P.~H.~J. Kelly, From h to p
  efficiently: Selecting the optimal spectral/hp discretisation in three
  dimensions, Mathematical Modelling of Natural Phenomena 6~(3) (2011) 84–96.
\newblock \href {http://dx.doi.org/10.1051/mmnp/20116304}
  {\path{doi:10.1051/mmnp/20116304}}.

\bibitem{Moxey16}
D.~Moxey, C.~Cantwell, R.~Kirby, S.~Sherwin,
  \href{https://www.sciencedirect.com/science/article/pii/S0045782516306739}{Optimising
  the performance of the spectral/hp element method with collective linear
  algebra operations}, Computer Methods in Applied Mechanics and Engineering
  310 (2016) 628--645.
\newblock \href {http://dx.doi.org/https://doi.org/10.1016/j.cma.2016.07.001}
  {\path{doi:https://doi.org/10.1016/j.cma.2016.07.001}}.
\newline\urlprefix\url{https://www.sciencedirect.com/science/article/pii/S0045782516306739}

\bibitem{Bolis14}
A.~Bolis, C.~D. Cantwell, R.~M. Kirby, S.~J. Sherwin,
  \href{https://onlinelibrary.wiley.com/doi/abs/10.1002/fld.3909}{From h to p
  efficiently: optimal implementation strategies for explicit time-dependent
  problems using the spectral/hp element method}, International Journal for
  Numerical Methods in Fluids 75~(8) (2014) 591--607.
\newblock \href
  {http://arxiv.org/abs/https://onlinelibrary.wiley.com/doi/pdf/10.1002/fld.3909}
  {\path{arXiv:https://onlinelibrary.wiley.com/doi/pdf/10.1002/fld.3909}},
  \href {http://dx.doi.org/https://doi.org/10.1002/fld.3909}
  {\path{doi:https://doi.org/10.1002/fld.3909}}.
\newline\urlprefix\url{https://onlinelibrary.wiley.com/doi/abs/10.1002/fld.3909}

\bibitem{Warburton19}
K.~Świrydowicz, N.~Chalmers, A.~Karakus, T.~Warburton,
  \href{https://doi.org/10.1177/1094342018816368}{Acceleration of
  tensor-product operations for high-order finite element methods}, The
  International Journal of High Performance Computing Applications 33~(4)
  (2019) 735--757.
\newblock \href {http://arxiv.org/abs/https://doi.org/10.1177/1094342018816368}
  {\path{arXiv:https://doi.org/10.1177/1094342018816368}}, \href
  {http://dx.doi.org/10.1177/1094342018816368}
  {\path{doi:10.1177/1094342018816368}}.
\newline\urlprefix\url{https://doi.org/10.1177/1094342018816368}

\bibitem{huynh}
H.~T. Huynh, A flux reconstruction approach to high-order schemes including
  discontinuous galerkin methods, in: 18th AIAA Computational Fluid Dynamics
  Conference, 2007.
\newblock \href {http://dx.doi.org/10.2514/6.2007-4079}
  {\path{doi:10.2514/6.2007-4079}}.

\bibitem{WITHERDEN2021113014}
F.~Witherden, P.~Vincent,
  \href{https://www.sciencedirect.com/science/article/pii/S0377042720303058}{On
  nodal point sets for flux reconstruction}, Journal of Computational and
  Applied Mathematics 381 (2021) 113014.
\newblock \href {http://dx.doi.org/https://doi.org/10.1016/j.cam.2020.113014}
  {\path{doi:https://doi.org/10.1016/j.cam.2020.113014}}.
\newline\urlprefix\url{https://www.sciencedirect.com/science/article/pii/S0377042720303058}

\bibitem{WITHERDEN20151232qq}
F.~Witherden, P.~Vincent,
  \href{https://www.sciencedirect.com/science/article/pii/S0898122115001224}{On
  the identification of symmetric quadrature rules for finite element methods},
  Computers \& Mathematics with Applications 69~(10) (2015) 1232--1241.
\newblock \href {http://dx.doi.org/https://doi.org/10.1016/j.camwa.2015.03.017}
  {\path{doi:https://doi.org/10.1016/j.camwa.2015.03.017}}.
\newline\urlprefix\url{https://www.sciencedirect.com/science/article/pii/S0898122115001224}

\bibitem{tetrasolp}
F.~Witherden, , J.~Park, P.~Vincent,
  \href{https://link.springer.com/article/10.1007/s10915-016-0204-y}{An
  analysis of solution point coordinates for flux reconstruction schemes on
  tetrahedral elements}, Journal of Scientific Computing 69~(2) (2016)
  905--920.
\newblock \href {http://dx.doi.org/https://doi.org/10.1007/s10915-016-0204-y}
  {\path{doi:https://doi.org/10.1007/s10915-016-0204-y}}.
\newline\urlprefix\url{https://link.springer.com/article/10.1007/s10915-016-0204-y}

\bibitem{hesthaven2007nodal}
J.~S. Hesthaven, T.~Warburton, Nodal discontinuous Galerkin methods:
  algorithms, analysis, and applications, Springer Science \& Business Media,
  2007.

\bibitem{CASTONGUAY2013400}
P.~Castonguay, D.~Williams, P.~Vincent, A.~Jameson,
  \href{https://www.sciencedirect.com/science/article/pii/S0045782513002156}{Energy
  stable flux reconstruction schemes for advection–diffusion problems},
  Computer Methods in Applied Mechanics and Engineering 267 (2013) 400--417.
\newblock \href {http://dx.doi.org/https://doi.org/10.1016/j.cma.2013.08.012}
  {\path{doi:https://doi.org/10.1016/j.cma.2013.08.012}}.
\newline\urlprefix\url{https://www.sciencedirect.com/science/article/pii/S0045782513002156}

\bibitem{vincent2011}
P.~Vincent, P.~Castonguay, A.~Jameson,
  \href{https://www.sciencedirect.com/science/article/pii/S0021999111004323}{Insights
  from von neumann analysis of high-order flux reconstruction schemes}, Journal
  of Computational Physics 230~(22) (2011) 8134--8154.
\newblock \href {http://dx.doi.org/https://doi.org/10.1016/j.jcp.2011.07.013}
  {\path{doi:https://doi.org/10.1016/j.jcp.2011.07.013}}.
\newline\urlprefix\url{https://www.sciencedirect.com/science/article/pii/S0021999111004323}

\bibitem{KoprivaDavidA1998ASMS}
D.~A. Kopriva, A staggered-grid multidomain spectral method for the
  compressible navier–stokes equations, Journal of computational physics
  143~(1) (1998) 125--158.

\bibitem{SunWangLiu2007}
Y.~Sun, Z.~Wang, Y.~Liu, {High-Order Multidomain Spectral Difference Method for
  the Navier-Stokes Equations on Unstructured Hexahedral Grids}, Communications
  in Computational Physics 2~(2) (2007) 310--333.

\bibitem{ToroRiemann}
E.~F. Toro, Riemann solvers and numerical methods for fluid dynamics a
  practical introduction, 3rd Edition, Springer, Berlin, 2006.

\bibitem{Rusanov}
V.~Rusanov, The calculation of the interaction of non-stationary shock waves
  and obstacles, U.S.S.R. computational mathematics and mathematical physics
  1~(2) (1962) 304--320.

\bibitem{xsmm}
A.~{Heinecke}, G.~{Henry}, M.~{Hutchinson}, H.~{Pabst}, {LIBXSMM}: Accelerating
  small matrix multiplications by runtime code generation, in: SC '16:
  Proceedings of the International Conference for High Performance Computing,
  Networking, Storage and Analysis, 2016, pp. 981--991.
\newblock \href {http://dx.doi.org/10.1109/SC.2016.83}
  {\path{doi:10.1109/SC.2016.83}}.

\bibitem{roofline}
S.~Williams, A.~Waterman, D.~Patterson,
  \href{https://doi.org/10.1145/1498765.1498785}{Roofline: An insightful visual
  performance model for multicore architectures}, Commun. ACM 52~(4) (2009)
  65–76.
\newblock \href {http://dx.doi.org/10.1145/1498765.1498785}
  {\path{doi:10.1145/1498765.1498785}}.
\newline\urlprefix\url{https://doi.org/10.1145/1498765.1498785}

\bibitem{VERMEIRE2017497}
B.~Vermeire, F.~Witherden, P.~Vincent,
  \href{https://www.sciencedirect.com/science/article/pii/S0021999116307136}{On
  the utility of gpu accelerated high-order methods for unsteady flow
  simulations: A comparison with industry-standard tools}, Journal of
  Computational Physics 334 (2017) 497--521.
\newblock \href {http://dx.doi.org/https://doi.org/10.1016/j.jcp.2016.12.049}
  {\path{doi:https://doi.org/10.1016/j.jcp.2016.12.049}}.
\newline\urlprefix\url{https://www.sciencedirect.com/science/article/pii/S0021999116307136}

\bibitem{HOReview}
Z.~Wang, K.~Fidkowski, R.~Abgrall, F.~Bassi, D.~Caraeni, A.~Cary, H.~Deconinck,
  R.~Hartmann, K.~Hillewaert, H.~Huynh, N.~Kroll, G.~May, P.-O. Persson, B.~van
  Leer, M.~Visbal,
  \href{https://onlinelibrary.wiley.com/doi/abs/10.1002/fld.3767}{{High-order
  CFD methods: current status and perspective}}, International Journal for
  Numerical Methods in Fluids 72~(8) (2013) 811--845.
\newblock \href
  {http://arxiv.org/abs/https://onlinelibrary.wiley.com/doi/pdf/10.1002/fld.3767}
  {\path{arXiv:https://onlinelibrary.wiley.com/doi/pdf/10.1002/fld.3767}},
  \href {http://dx.doi.org/https://doi.org/10.1002/fld.3767}
  {\path{doi:https://doi.org/10.1002/fld.3767}}.
\newline\urlprefix\url{https://onlinelibrary.wiley.com/doi/abs/10.1002/fld.3767}

\bibitem{VANREES20112794}
W.~M. {van Rees}, A.~Leonard, D.~Pullin, P.~Koumoutsakos,
  \href{https://www.sciencedirect.com/science/article/pii/S0021999110006467}{A
  comparison of vortex and pseudo-spectral methods for the simulation of
  periodic vortical flows at high reynolds numbers}, Journal of Computational
  Physics 230~(8) (2011) 2794--2805.
\newblock \href {http://dx.doi.org/https://doi.org/10.1016/j.jcp.2010.11.031}
  {\path{doi:https://doi.org/10.1016/j.jcp.2010.11.031}}.
\newline\urlprefix\url{https://www.sciencedirect.com/science/article/pii/S0021999110006467}

\bibitem{qcrit}
J.~{Hunt}, A.~{Wray}, P.~{Moin}, Eddies, streams, and convergence zones in
  turbulent flows, in: Proceedings of the Summer Program, Center for Turbulence
  Research, Stanford Univ., 1998, pp. 193--208.

\bibitem{ZEFR}
J.~Romero, J.~Crabill, J.~Watkins, F.~Witherden, A.~Jameson,
  \href{https://www.sciencedirect.com/science/article/pii/S0010465520300229}{{ZEFR}:
  A {GPU}-accelerated high-order solver for compressible viscous flows using
  the flux reconstruction method}, Computer Physics Communications 250 (2020)
  107169.
\newblock \href {http://dx.doi.org/https://doi.org/10.1016/j.cpc.2020.107169}
  {\path{doi:https://doi.org/10.1016/j.cpc.2020.107169}}.
\newline\urlprefix\url{https://www.sciencedirect.com/science/article/pii/S0010465520300229}

\end{thebibliography}

\end{document}